\documentclass[usenatbib]{mnras}

\usepackage{newtxtext,newtxmath}

\usepackage[T1]{fontenc}
\usepackage{ae,aecompl}

\usepackage{graphicx}	
\usepackage{amsmath}	

\usepackage{amssymb}	

\newcommand{\blindcorrection}[1]{#1}

\title[Atomic species inventory in WASP-121b]{An inventory of atomic species in the atmosphere of WASP-121b using UVES high-resolution spectroscopy}

\author[S. R. Merritt et al.]{
Stephanie R. Merritt$^{1}$\thanks{E-mail: smerritt01@qub.ac.uk},
Neale P. Gibson$^{2}$,
Stevanus K. Nugroho$^{1}$,
\newauthor Ernst J. W. de Mooij$^{1}$,
Matthew J. Hooton$^{3}$,
Joshua D. Lothringer$^{4}$, 
\newauthor Shannon M. Matthews$^{1}$, 
Thomas Mikal-Evans$^{5}$,
Nikolay Nikolov$^{6}$,
David K. Sing$^{4, 7}$,
\newauthor and Chris A. Watson$^{1}$
\\
$^{1}$Astrophysics Research Centre, School of Mathematics and Physics, Queen's University Belfast, Belfast BT7 1NN, UK\\
$^{2}$School of Physics, Trinity College Dublin, Dublin 2, Ireland\\
$^{3}$Physikalisches Institut, Universität Bern, Gesellschaftsstrasse 6, 3012, Bern, Switzerland\\
$^{4}$Department of Physics and Astronomy, Johns Hopkins University, Baltimore, MD, USA\\
$^{5}$Kavli Institute for Astrophysics and Space Research, Massachusetts Institute of Technology, 77 Massachusetts Avenue, 37-241, Cambridge, MA 02139, USA\\
$^{6}$Space Telescope Science Institute, 3700 San Martin Dr, Baltimore, MD 21218, USA\\
$^{7}$Department of Earth and Planetary Sciences, Johns Hopkins University, Baltimore, MD, USA\\
}

\date{Accepted XXX. Received YYY; in original form ZZZ}

\pubyear{2020}

\begin{document}
\label{firstpage}
\pagerange{\pageref{firstpage}--\pageref{lastpage}}
\maketitle

\begin{abstract}
Ultra-hot Jupiters (UHJs) present excellent targets for atmospheric characterisation. Their hot dayside temperatures (T $\gtrsim$ 2200 K) strongly suppress the formation of condensates, leading to clear and highly-inflated atmospheres extremely conducive to transmission spectroscopy. Recent studies using optical high-resolution spectra have discovered a plethora of neutral and ionised atomic species in UHJs, placing constraints on their atmospheric structure and composition. Our recent work has presented a search for molecular features and detection of \ion{Fe}{i} in the UHJ WASP-121b using VLT/UVES transmission spectroscopy. Here, we present a systematic search for atomic species in its atmosphere using cross-correlation methods. In a single transit, we uncover \blindcorrection{potential signals of 17 atomic species which we investigate further, categorising 5 as strong detections, 3 as tentative detections, and 9 as weak signals worthy of further exploration. We confirm previous detections of \ion{Cr}{i}, \ion{V}{i}, \ion{Ca}{i}, \ion{K}{i} and exospheric \ion{H}{i} and \ion{Ca}{ii} made with HARPS and ESPRESSO, and independently re-recover our previous detection of \ion{Fe}{i} at 8.8\,$\sigma$ using both the blue and red arms of the UVES data. We also add a novel detection of \ion{Sc}{ii} at 4.2\,$\sigma$.} Our results further demonstrate the richness of UHJs for optical high-resolution spectroscopy.
\end{abstract}




\section{Introduction}
\label{sec:intro}
The recently-emerging class of exoplanets known as ultra-hot Jupiters (hereafter UHJs) present an intriguing subject for study and characterisation. These tidally-locked gas giants orbit on extremely short periods around their parent stars and thus experience extreme irradiation, increasing their dayside temperatures to the point where their chemistry and atmospheric structure are expected to differ greatly from cooler hot Jupiters (dayside T $\gtrsim$ 2200 K, e.g. \citealt{Parmentier2018}). The increased temperature leads to the dissociation of molecules and the partial thermal ionisation of atomic species in the hot dayside, thereby preventing the formation of the high-altitude cloud decks and aerosol particles that dominate the spectra of their cooler siblings \citep{Helling2019}. Instead, the dominant source of continuum opacity comes from scattering by H$^-$ ions created by the dissociation of molecular hydrogen and the abundance of free electrons from the thermal ionisation of metals \citep{Arcangeli2018}. The dissociation of molecular hydrogen at temperatures over 2500 -- 3000 K also leads to atomic hydrogen becoming the dominant constituent of the atmosphere \citep{Lothringer2018, Parmentier2018, Kitzmann2018}, reducing the mean molecular weight of the atmosphere. The effects of the dissociation of molecular hydrogen, and its recombination on the cooler night-side, is also theorised to lead to increased day-side/night-side heat transport in UHJs \citep{BellCowan2018}, further altering our expectations of the atmospheric chemistry from the assumptions made for cooler hot Jupiters. 

These combined characteristics make UHJs excellent targets for atmospheric characterisation via their transmission spectra (the fingerprint of the exoplanet spectrum found as the light from the host star passes through the upper layers of the atmosphere during transit). The reduced mean molecular weight of the atmosphere and high temperature increases the scale height of the atmosphere, extending the amount of atmosphere observable during transit events. The lack of high-altitude clouds and aerosols ensures that atomic and molecular features are less suppressed by these sources of opacity, as can often happen in cooler hot Jupiters \citep[e.g.][]{Gibson2013a, Gibson2013b, Wakeford2017, Kirk2017, May2018, Espinoza2019, Wilson2020}. Additionally, when the terminator of the atmosphere is probed in transit observations, multiple temperature regimes are explored in a transition from the hot day-side to the cooler night-side \citep{Parmentier2018}, where both ionised and neutral metals/atoms and recombined molecules may be observable at the same time. Simulations of UHJs using stellar models have predicted a plethora of neutral and ionised atomic species in their atmospheres \citep{Lothringer2018, Lothringer2020, LothringerBarman2019}.

The adaptation of the high-resolution Doppler spectroscopy technique \citep{Snellen2010}, originally used to characterise spectroscopic binary systems, \blindcorrection{has provided us with a new tool to inventory the atomic and molecular constituents of exoplanet atmospheres. The large radial velocity of the planet imparts a correspondingly significant Doppler shift in its atomic and molecular lines. \blindcorrection{The planetary signal can then be separated from the telluric and stellar lines in the spectra by using detrending techniques to strip all static and quasi-static trends from the time series. The resulting residuals contain the photon noise and, hidden within, the Doppler-shifted lines from the planet's transmission spectrum. As these lines are individually resolved at high resolution, they can be extracted from the noise by cross-correlation with template spectra} of the atomic or molecular species of interest, effectively summing up over hundreds or thousands of individually-resolved spectral lines and strengthening the detection signal.}

\blindcorrection{This method has, until recently, seen most of its success in the detection of molecules at infrared wavelengths \citep[e.g][]{Brogi2012, Birkby2013, Lockwood2014}. However, Doppler spectroscopy at optical wavelengths has recently played a huge part in the characterisation of ultra-hot Jupiters.} This technique was used in a highly-successful attempt to inventory the species present in the atmosphere of possibly the most notorious UHJ of them all, KELT-9b. The hottest-known exoplanet to date with a $T_{eq}$ of 4050 K, \citet{Hoeijmakers2018, Hoeijmakers2019} revealed the presence of \ion{Mg}{i}, \ion{Fe}{i}, \ion{Fe}{ii}, \ion{Ti}{ii}, \ion{Na}{ii}, \ion{Na}{i}, \ion{Cr}{ii}, \ion{Sc}{ii} and \ion{Y}{ii}, along with hints of \ion{Ca}{i}, \ion{Cr}{i}, \ion{Co}{i} and \ion{Sr}{ii}. High-resolution spectroscopy has also provided detections of \ion{Fe}{i}, \ion{Fe}{ii}, \ion{Ca}{i}, \ion{Ca}{ii}, \ion{Mg}{i}, \ion{Cr}{ii} and the Balmer series of hydrogen in the UHJ KELT-20b/MASCARA-2b \citep{CasasayasBarris2018, CasasayasBarris2019, Nugroho2020_k20b, Hoeijmakers2020_kelt20b, Stangret2020}. High-resolution explorations of the UHJ WASP-33b revealed signs of TiO \citep{Nugroho2017}, \ion{Fe}{i} \citep{Nugroho2020_w33b}, and \ion{Ca}{ii} \citep{Yan2019}, though TiO was not detected in later observations by \citet{Herman2020}. A search for FeH in several hot Jupiters resulted in tentative evidence for its presence in WASP-33b and KELT-20b/MASCARA-2b \citep{Kesseli2020}. \blindcorrection{An asymmetrical absorption signal of iron during transit of WASP-76b was attributed by \citet{Ehrenreich2020} to the existence of a chemical gradient.}

One especially interesting target for atmospheric characterisation, whether at high-resolution or low, is the UHJ WASP-121b \citep{Delrez2016}.\blindcorrection{ WASP-121b is in orbit around a bright (V = 10.5) F6V-type star \citep{Hoeg2000}, and its period of just 1.27 days imparts an equilibrium temperature of 2400 K. Low-resolution observations with HST have resulted in a detection of water in transmission, with tentative evidence for TiO or VO \citep{Evans2016, Evans2018, Tsiaras2018}; water was also detected in emission, leading to the first direct measurement of a temperature inversion} \citep{Evans2017}. This temperature inversion was later confirmed by phase-curve photometry from TESS (the Transiting Exoplanet Survey Satellite;  \citealt{Daylan2019, Bourrier2019}). New secondary eclipse measurements made by HST recently reconfirmed the presence of water emission features \citep{Mikal-Evans2020}, though these observations have failed to reproduce a feature in the emission spectrum previously attributed to VO  in \citet{Evans2017} and \citet{Mikal-Evans2019}, since attributed to systematics. Recent UV observations taken during transit, also with HST, \blindcorrection{discovered \ion{Fe}{ii} and \ion{Mg}{ii} features extending far higher in the atmosphere than previously-detected features at redder wavelengths, evidence that WASP-121b has an extended, escaping atmosphere \citep{Sing2019}. A similar excess in UV absorption had previously been detected by \citet{Salz2019}. Eclipse observations taken in the $z^{\prime}$ band by \citet{Mallonn2019} placed upper limits on WASP-121b's albedo, and} \blindcorrection{potential variability in its atmosphere has been posited based on Gemini/GMOS observations \citep{Wilson2021}.}

At high resolution, a search for TiO or VO in the transmission spectrum of WASP-121b using UVES (UV-Visual Echelle Spectrograph) on the VLT failed to detect either \citep{Merritt2020}, in conflict with earlier low-resolution work \citep{Evans2016, Evans2018, Tsiaras2018}, though the inaccuracy of the VO high-temperature line-list was cited as a possible cause for the non-detection of VO. Analysis by \citet{Gibson2020} of the same UVES transits, and by \citet{Bourrier2020} and \citet{Cabot2020} using HARPS (High-Accuracy Radial-velocity Planet Searcher) on La Silla, have \blindcorrection{detected} \ion{Fe}{i} deeper in the atmosphere, with \citet{Cabot2020} additionally detecting the presence of H$\alpha$. Analysis of the same HARPS data by \citet{Ben-Yami2020} confirmed the presence of \ion{Fe}{i} and \ion{Fe}{ii}, while presenting additional detections of \ion{V}{i} and \ion{Cr}{i} and a non-detection of \ion{Ti}{i}. Further analysis by \citet{Hoeijmakers2020}, again on the same HARPS observations, confirmed the presence of \ion{Fe}{i}, \ion{V}{i} and \ion{Cr}{i}, while adding detections of \ion{Mg}{i}, \ion{Na}{i}, \ion{Ca}{i} and \ion{Ni}{i}, and confirming non-detections of \ion{Ti}{i} and TiO. \blindcorrection{Finally, a study by \citet{Borsa2021} using data from ESPRESSO/VLT confirmed many of the previous detections and added novel detections of \ion{K}{i}, \ion{Li}{i} and exospheric \ion{Ca}{ii}.}  This wealth of information from only a handful of datasets showcases the huge potential of high-resolution spectroscopy as a tool to explore the atmospheres of UHJs.

In order to confirm and expand upon previous detections, we present an atomic species inventory of the \blindcorrection{ultra-hot Jupiter WASP-121b, using high-resolution spectra taken with VLT/UVES \citep{Dekker2000}, an instrument which has seen successful use for the exploration of exoplanet atmospheres \citep{Snellen2004, Khalafinejad2017, Gibson2019}. The blue arm of these observations was previously explored by \citep{Gibson2020}}, and analysis of the red arm in a search for molecular features was presented in \citet{Merritt2020}: this paper presents analysis of both the red and blue arms combined with the aim of searching for a broad range of species. In Sec.~\ref{sec:obs}, we describe the observations and the extraction of the spectra. Sec.~\ref{sec:analysis} discusses our data-processing steps, the creation of model transmission spectra for cross-correlation, the cross-correlation process and our injection tests. Results of a broad-scale atomic species search are presented alongside our detection criteria in Sec.~\ref{sec:res}; the results are further discussed in detail in Sec.~\ref{sec:dis}, including the results of our injection tests and notable non-detections.

\section{Observations And Data Reduction}
\label{sec:obs}
Two transits of WASP-121b were observed using the high-resolution echelle spectrograph UVES, mounted on the 8.2m `Kueyen' telescope (UT2) of the VLT. Observations were taken on the nights of 25 December 2016 and 4 January 2017 as part of program 098.C- 0547 (PI: Gibson). The second transit was discarded due to low signal-to-noise caused by a loss in guiding. The peak S/N over the course of the observations can be found in \citet{Merritt2020}. The blue arm of this data was previously presented in \citet{Gibson2020}, and the red arm of this data was previously presented in \citet{Merritt2020}: further details on the observations can be found there. The subsequent analysis in this paper uses both arms for a total of 64 spectral orders from 375 nm to 866 nm.

Extraction proceeded almost identically to the process outlined in \citet{Merritt2020}, using the custom \texttt{Python} pipeline outlined in that work. However, bias subtraction was found to lead to spuriously negative values in low-flux areas of the spectrum in the blue arm, and a correction from the overscan areas was implemented. Additionally, the order traces from the esorex pipeline were found to be slightly off-centre, and were corrected. \blindcorrection{Once again, the five spectral traces filled the extraction aperture almost completely, so no background subtraction could be performed. Calculation of the barycentric Julian date BJD$_{\text{TDB}}$ for each observation was performed using functions from the \texttt{astropy.time} \texttt{Python} package.}

Outliers were removed from the spectra by a two-stage process slightly improved from the method outlined in \citet{Merritt2020}. First, the spectra were sigma-clipped for values over 3 sigma, which were replaced by the median value of the surrounding 100 pixels. \blindcorrection{As this did not remove all outliers,} a single-component PCA reconstruction of the time-series spectra then was subtracted from the spectra to remove all time- and wavelength-dependent trends. Two iterations of sigma-clipping were performed upon the residuals, with the values of any outliers set to zero: \blindcorrection{adding the PCA reconstruction back to the data then replaced the clipped outliers with their value from the PCA reconstruction.} \blindcorrection{Although a 3\,$\sigma$ clipping might appear aggressive, it replaced an average of only 0.07\% of pixels per order, and does not impact our results.}

In an additional divergence from \citet{Merritt2020}, spectral alignment was performed to correct for small shifts in the original wavelength solution provided by the \texttt{esorex} pipeline. This was not thought to be necessary in the search for TiO and VO presented in \citet{Merritt2020}, as the total wavelength drift over each night was found to be significantly less than a resolution element ($\sim$2.5 $\text{km s}^{-1}$ at R $\sim$ 110,000). However, \citet{Gibson2020} found that order-to-order drift in the blue arm was  $\sim$3 $\text{km s}^{-1}$, probably due to inaccuracies in the wavelength solution provided by the pipeline. Additionally, as the goal of this work is to search for atomic species, many of which may be present in the stellar spectrum, the need for an accurate wavelength solution is increased in order to optimise the detrending methods used to remove the stellar spectrum from the data, as outlined in Sec.~\ref{sec:prepro}.

The blue arm of the spectra was aligned via cross-correlation with a PHOENIX model for an F6V-type star \citep{Husser2013}. \blindcorrection{Though this would also fit for the Rossiter-McLaughlin effect, WASP-121b is on a near-polar orbit with a projected obliquity of $258^{\circ}$ \citep{Delrez2016}, and the RM effect is expected to have an amplitude well below both the change in velocity of WASP-121b during transit and the resolution limit of UVES \citep{Gibson2020}. The RM effect is therefore expected to have a negligible effect on both alignment and on species detection, with the latter confirmed via modelling by \citet{Cabot2020}.}

The resultant detected velocity shifts were smoothed with a high-pass filter and applied to each frame \blindcorrection{(via linear interpolation)} and order of the blue arm data separately. In the red arm, alignment could not be successfully performed to the stellar spectrum due to the lack of lines and the contamination due to the presence of tellurics. Instead, alignment was performed via cross-correlation to a telluric template generated from ESO's SkyCalc tool \citep{Noll2012, Jones2013}. In the red arm, the order-to-order discrepancy in the wavelength solution was found to be much less ($\sim$1 $\text{km s}^{-1}$), and due to the sporadic placement of tellurics throughout the red arm spectra, order-specific alignment was very difficult for orders with few tellurics. As a result, the wavelength corrections for each frame were averaged over the orders and applied on a frame-by-frame basis only.

\blindcorrection{
We note that in the process used to correct the wavelength solution, the spectra in the blue arm are aligned to the stellar rest frame. This simultaneously corrects for the time-dependent barycentric velocity variations and the systemic velocity of WASP-121. As the spectra in the red arms are aligned to the telluric rest frame they are not yet corrected for these effects. We later re-align the red arm to the stellar rest frame; however, as the tellurics dominate the systematics in the red arm (which is also the reason why telluric features were used for the alignment), we first perform systematic corrections before performing the re-alignment. This is discussed in Sec.~\ref{sec:prepro}.
}
\blindcorrection{Simultaneously with the correction of the wavelength solution, the spectra were also linearised in wavelength and super-sampled to 2/3 of the average pixel width for each order.} Minimising the number of interpolations performed on the data helps preserve spectral information and reduces the introduction of noise. The results of the extraction were cross-checked with an independent pipeline \citep{Gibson2020}.

\section{Analysis}
\label{sec:analysis}
\subsection{Pre-processing}
\label{sec:prepro}
\begin{figure*}
	\includegraphics[width=\textwidth]{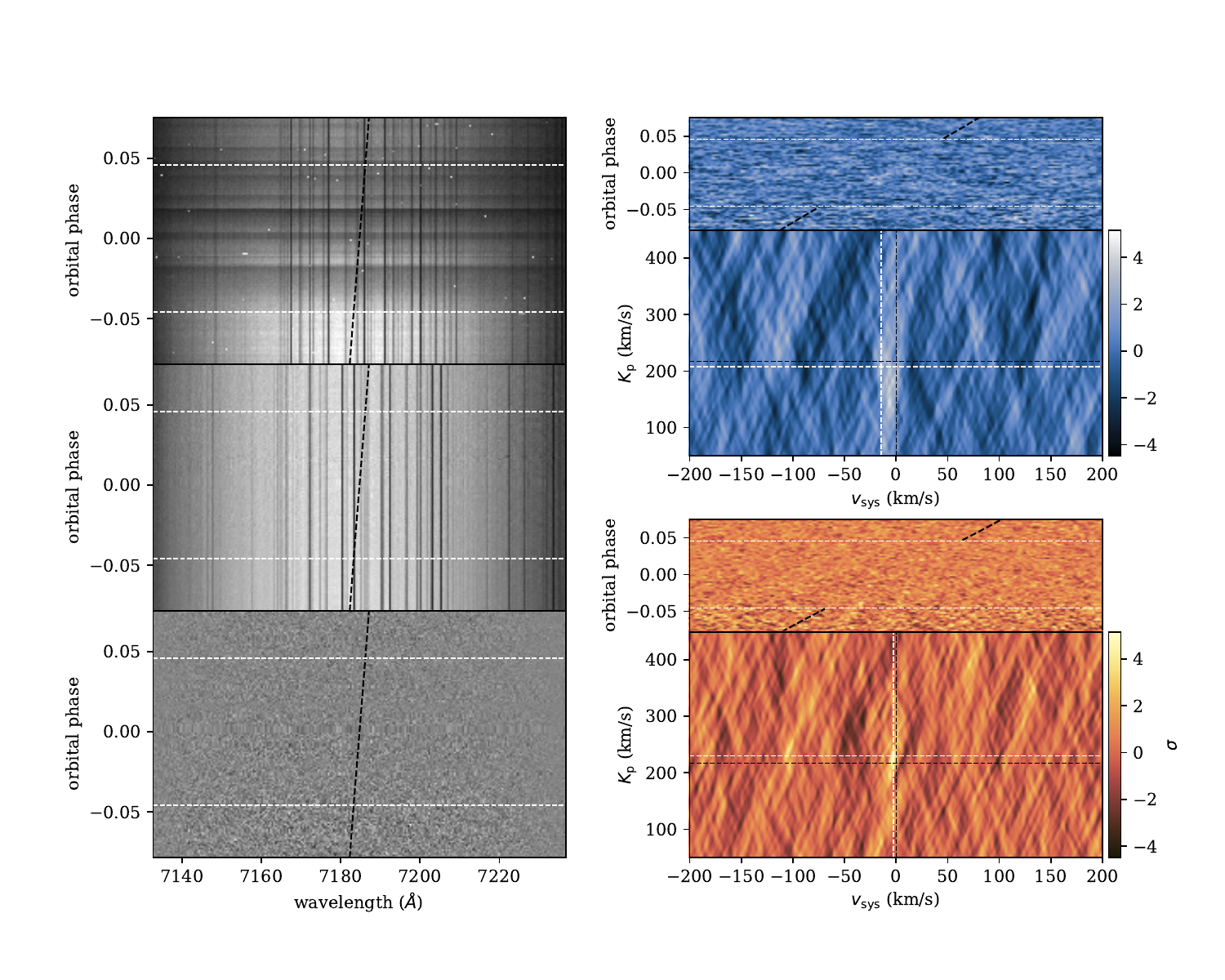}
    \caption{Pre-processing and cross-correlation. Left: Data processing of a single raw echelle order (top), after wavelength alignment and correction for blaze variation (middle), and after division by \textsc{SysRem} model and weighting by uncertainties (bottom). The dashed white lines indicate the times of ingress and egress. The dashed black line indicates the approximate velocity shift of WASP-121b. Right: the summed cross-correlation function time series (upper) and cross-correlation velocity map (lower) for the blue (top) and red (bottom) arms of the data respectively, for the \ion{Cr}{i} model used to retrieve the signal presented in Fig~\ref{fig:majorresults}. The white dashed lines in the CCF time series plots represent the times of ingress and egress. The black dashed lines indicate the expected position of the signal `trace' in the CCF time series. In the velocity maps, the black dashed cross-hairs indicate the expected position of the signal; the white dashed cross-hairs mark the peak of the detected signal.}
    \label{fig:prepro}
\end{figure*}

We follow an almost identical methodology to that outlined in \citet{Merritt2020}, based on techniques \blindcorrection{often used in searches for molecular and atomic species in high-resolution spectra} \citep[e.g.][]{Snellen2010, Brogi2012, Birkby2013, Nugroho2020_k20b}. In order to search for the buried exoplanetary spectral signal, we must first remove all other trends from the time-series which are static or quasi-static in time, including the stellar spectrum and telluric absorption from the Earth's atmosphere, further modulated by the (time-varying) response of the instrument. The removal of these trends should, in theory, result in residuals which are composed of purely photon noise and the Doppler-shifted, continuum-removed planetary signal. In practice however, the pre-processing is imperfect, and neither perfectly removes the stellar and telluric lines, nor leaves the atmospheric signature of the planet untouched. Here, we briefly recap the data processing, and an example on a single spectral order is shown in Fig.~\ref{fig:prepro}.

First, we place all of the spectra on a `common' blaze function, which can vary with time. The correction is performed as in \citet{Merritt2020} by dividing each order of time-series spectra by the median spectrum, smoothing the resulting residual spectra with a median filter with a width of 15 pixels and a Gaussian filter with a standard deviation of 50 pixels, and dividing through each by its smoothed residual spectra to correct the blaze variation.

Next, all static and quasi-static trends in time -- such as the `common' blaze function, the stellar spectrum and the telluric spectrum -- are \blindcorrection{removed from the spectral time-series using the \textsc{SysRem} algorithm \citep{Tamuz2005}, a tool in standard use for the de-trending of high-resolution time-series data \citep[e.g.][]{Birkby2013, Birkby2017, Nugroho2020_k20b}. One advantage \textsc{SysRem} has over other PCA-based detrending algorithms is its inherent treatment of the pixel uncertainties.} Unlike \citet{Merritt2020}, we use the Poisson pixel uncertainties calculated from the extracted spectra, which resulted in more efficient removal of the stellar signal than the outer-product variance method described in that work. 

The blaze correction was \blindcorrection{found to be unstable at the bluest end of each spectral order, an effect seen in both arms of the data}. We thus remove the first 900 and last 90 pixels of each order in the blue arm (identically to  \citealt{Gibson2020} after accounting for 3/2 supersampling) and the first 750 pixels of the red arm (identically to \citealt{Merritt2020}, likewise). This removes 22\% of the pixels in the blue arm spectra and 12\% in the red arm. While this is a significant chunk of the data, the edges of the orders are low in S/N, therefore this has minimal impact on our detections.

We run multiple passes of \textsc{SysRem} to create a 2D model representation for each time-series. Instead of using the resulting model-subtracted residuals as our final product, we instead sum the \textsc{SysRem} models for each pass and divide our time-series spectra by this model before subtracting 1. This process preserves the relative line strengths of the planet's transmission spectrum, without first having to divide through by the stellar spectrum \citep{Gibson2020}. The data uncertainties are also divided through by the model.
The detrended residuals are then optimally weighted by the square of the pixel Poisson uncertainties.

In the blue arm, where stellar lines dominate, \textsc{SysRem} is performed in the rest frame of the star. In the red arm, \textsc{SysRem} is performed in the telluric rest frame, in order to optimise the removal of the dominant telluric spectrum.
Thus, after \textsc{SysRem} detrending, the red arm is then further corrected for the systemic and barycentric velocities as outlined in Sec.~\ref{sec:obs}, to bring the red arm spectra into the stellar rest frame. \blindcorrection{We note that the barycentric correction and systemic velocity are known to much higher precision than the instrument resolution, and therefore the different methods used to align the spectra in the red and blue arms will not impact our results.}

\subsection{Model transmission spectra}
\label{sec:models}

\begin{figure*}
	\includegraphics[width=\textwidth]{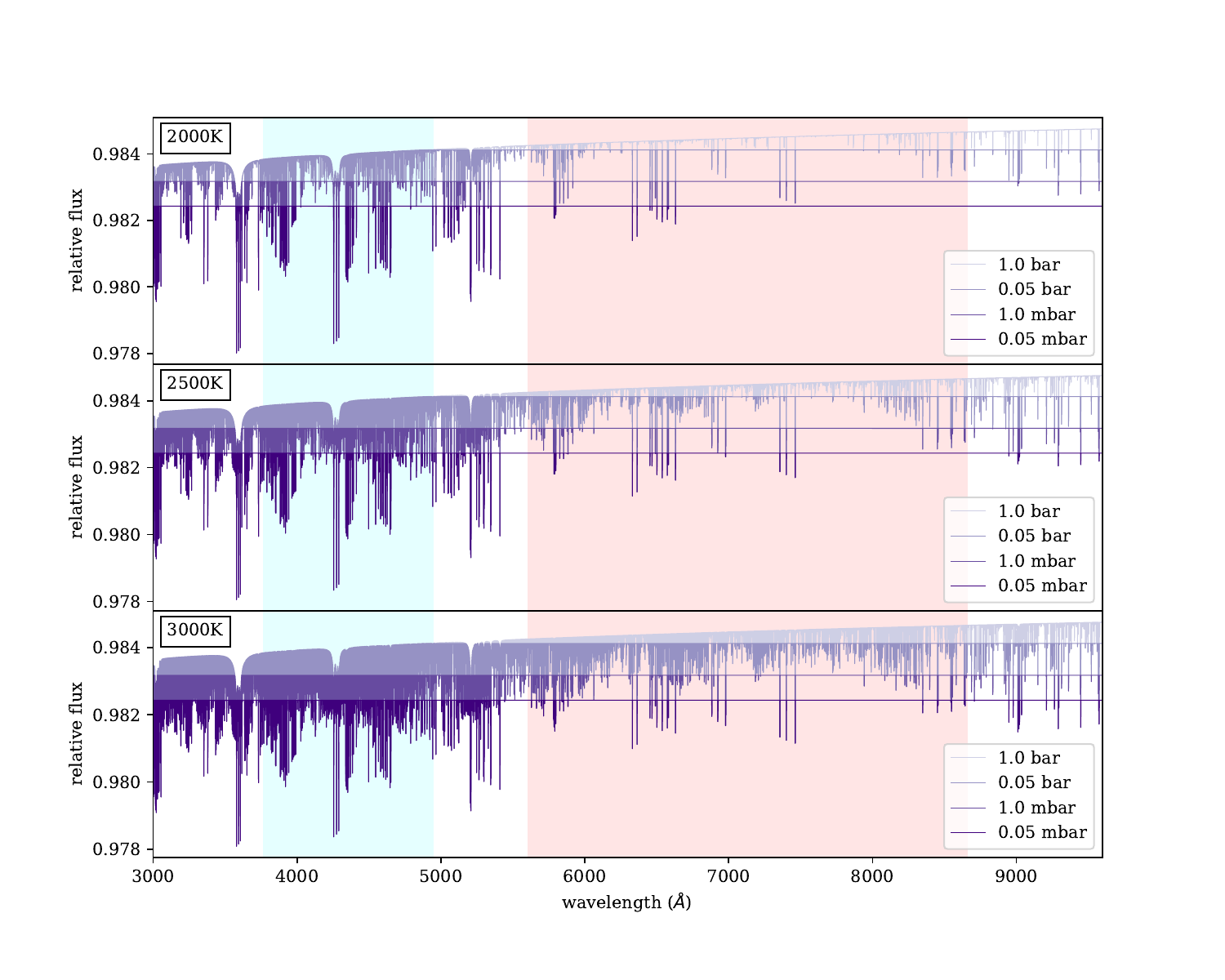}
    \caption{Examples of the model transmission spectra for \ion{Cr}{i} in WASP-121b at three different temperatures and four different positions of the grey cloud deck $P_{\text{cloud}}$. The log volume mixing ratio is held at $10^{-8}$. The blue and red regions indicate the blue and red wavelength ranges of the UVES data.}
    \label{fig:modelex}
\end{figure*}

\blindcorrection{Detecting atomic features using the cross-correlation technique} requires spectral models of each species. \blindcorrection{To create these models, high-temperature line-lists of the species of interest are used to generate absorption cross-sections. These cross-sections are then incorporated into radiative transfer models, which generate model transmission spectra for cross-correlation.}

Cross-sections for neutral atomic species and their singly-ionised counterparts numbering from atomic number one -- hydrogen -- to atomic number 39 -- yttrium -- were generated following \citet{Nugroho2017, Nugroho2020_k20b, Nugroho2020_w33b} using \texttt{HELIOS-K} \citep{GrimmHeng2015} with the Kurucz atomic line lists \citep{Kurucz2018}. We used a Voigt line profile with thermal and natural broadening only at a resolution of 0.01 cm$^{-1}$ with an absolute line wing cut-off of $1 \times 10^{7}$ cm$^{-1}$, except for \ion{H}{i} and \ion{He}{i} which used $3 \times 10^{4}$ cm$^{-1}$ and $1 \times 10^{5}$ cm$^{-1}$ respectively. These values were chosen to ensure important lines are not washed away. We chose to explore no further than yttrium (as beyond this point, solar abundances become very low) with the exception of \ion{Cs}{i}, due to its presence in late-M and L-dwarfs \citep[e.g.][]{Basri2000}.

Rather than producing a transmission spectrum for each species via numerical integration over a layered atmosphere, we instead elect to use the analytical equation for an approximate transmission spectrum used in \citet{Gibson2020}, adapted from the work of \citet{Heng2017}. This equation computes the effective radius as:

\[
R(\lambda) = R_0 + H \left [ \gamma + \ln \left ( \frac{P_0}{mg}\sqrt{\frac{2\pi R_0}{H}} \right ) \right ] + H \ln \sum_{j} \chi_j \sigma_j(\lambda).
\]

\noindent Here, $H$, $g$, and $m$ are the scale height, surface gravity and mean molecular mass of the atmosphere; $R_0$ and $P_0$ are the reference radius and pressure, $\gamma = 0.56$ is a dimensionless constant, $\chi_j$ is the volume mixing ratio of species $j$, and $\sigma_j$ is the cross-section of species $j$. Scattering is accounted for by including Rayleigh scattering using the cross-section of \citet{LecavelierDesEtangs2008} and a grey cloud deck simulated by simply truncating the model at a given pressure level $P_{\text{cloud}}$. We disregard the important effects of H$^{-}$ scattering \citep{Parmentier2018} in order \blindcorrection{to avoid any further degeneracy with Rayleigh scattering and the cloud deck}.

\citet{Gibson2020} found this to produce models of sufficient accuracy for high-resolution detections, by comparing directly with a full radiative transfer model, as long as the same common assumptions of no pressure-broadening, a well-mixed atmosphere, and an isothermal atmosphere are made. The main benefit of this analytical equation lies in its speed, which allows us to swiftly create hundreds of spectra for a range of parameters. This greatly increases the efficiency of a broad species search and allows quick exploration of the effects of changing parameters with minimal loss in the accuracy of the template.

\begin{table}
\centering
\begin{tabular}{cc}
Parameter						& Value 								\\
\hline \hline
																\\
\textbf{WASP-121} 				& 									\\
\hline
																\\
$M_{\star}~(M_{\odot})$			& $1.353^{+0.080}_{-0.079}{~}^\text{a}$			\\
$R_{\star}~(R_{\odot})$			& $1.458 \pm 0.080{~}^\text{a}$				\\
Spectral type 					& F6V${~}^\text{a}$ 						\\
$T_{\textrm{eff}}$ (K) 			& $6459 \pm 140{~}^\text{a}$					\\
$V$-magnitude					& $10.44{~}^\text{a}$ 						\\
$v_{\textrm{sys}}$ ($\text{km s}^{-1}$) 				& $38.36 \pm 0.43{~}^\text{b}$					\\
							&									\\
\textbf{WASP-121b} 				&									\\
\hline
																\\
$T_{0}~(\textrm{BJD}_{(TDB)})$ 	& $2457599.551478 \pm 0.000049{~}^\text{c}$	\\
$P$ (days)					& $1.2749247646 \pm 0.0000000714{~}^\text{c}$	\\
$a/R_{\star}$ 					& $3.86  \pm 0.02{~}^\text{d}$					\\
$R_{p}/R_{\star}$				& $0.1218 \pm 0.0004{~}^\text{d}$				\\
$i~(\deg)$						& $89.1 \pm 0.5{~}^\text{d}$					\\
$T_{\textrm{eq}}$ (K)			& $> 2400$							\\
$H$ (km)						& $\sim 960^\text{d}$							\\
$K_p$ ($\text{km s}^{-1}$)                    & $\sim 217 $ \\  \\ \hline
\\ 
\end{tabular}
\caption{Stellar and planetary parameters for the WASP-121b system utilised in this paper. Values marked with (a) are adopted from \citet{Delrez2016}; (b) from \citet{Gaia2018}; (c) from \citet{Sing2019}; and (d) from \citet{Evans2018}.}
\label{table:param}
\end{table}

We generated spectra for every species at three temperatures of 2000 K, 2500 K and 3000 K based on the most recent estimates of the temperature of WASP-121b at the limb \citep{Mikal-Evans2019, Gibson2020}. We additionally generated models using a higher set of temperatures (from 3500 K to 9500 K, in steps of 500 K) for the seven ions for which we found signals (given in Sec.~\ref{sec:res}). We assume a Jupiter-like composition for the mean molecular mass, and the scale height, which is expected to vary with temperature, was instead fixed at H = 960 km using a temperature of 2800 K (approximately the temperature at the top of the atmosphere as measured by \citealt{Mikal-Evans2019}). As we are subtracting the continuum from both the data (as outlined in Sec.~\ref{sec:prepro}) and the model spectra themselves, we are not particularly sensitive to the scale height as a parameter: changing the scale height simply scales the model amplitude, which should not affect the results of the cross-correlation detections (although does impact the injection tests). The reference radius and pressure were set at 1.8 $R_J$ and 20 mbar respectively. A full list of system and planetary parameters used in this work is presented in Table~\ref{table:param}.

Instead of varying the volume mixing ratio or VMR of each species, as was done in \citet{Merritt2020}, we choose to vary the position of the grey cloud deck, $P_{\text{cloud}}$. Four values of $P_{\text{cloud}}$ were chosen (1 bar, 0.05 bar, 1 mbar, 0.05 mbar). These were not physically motivated but instead selected to vary the amount of truncation of each model, to give a good range of lines present above the continuum, for an arbitrarily-chosen VMR of $10^{-8}$. An example of the 12 models created for a single species (\ion{Cr}{i}) is shown in Fig.~\ref{fig:modelex}. Given the degeneracy between reference pressure, radius and abundance \citep[e.g.][]{Benneke2012, Heng2017}, this approach is not sensitive to the abundances. Instead, this forces us to consider detections entirely in the sense of line strength present above the continuum (the only parameter we can meaningfully measure with our methodology).

This process results in 1312 model spectra being generated: three temperatures and four values of $P_{\text{cloud}}$ for 40 neutral species and 37 singly-ionised forms, plus an additional 13 temperatures for seven of the ionised species. However, we then eliminate any models with no or extremely weak lines in the wavelength ranges of the blue and red arms of our data, reducing the number to 792 models covering 43 neutral and ionised species. We did not constrain our model selection in any other way, such as by the supposed detectability of the species, as such constraints are invariably model-dependent.

Finally, before cross-correlation, the continuum and scattering profile is removed from all of the model spectra, as the pre-processing steps outlined in Sec.~\ref{sec:prepro} remove all large-scale variation from our data, including from any extant exoplanetary signal. This continuum-removal is performed by subtracting a low-pass filter consisting of a 1400-pixel maximum filter applied in the wavelength direction which is subsequently smoothed by a Gaussian filter with a kernel of 400 pixels, removing all large-scale variations in the model and normalising it at zero.

\subsection{Cross-correlation}
\label{sec:cc}
As described in Sec.~\ref{sec:prepro}, we divide the spectra (and uncertainties) through by a \textsc{SysRem} model in order to remove all dominant trends before weighting the residuals by the pixel variance. We create three sets of weighted \textsc{SysRem} residuals for each arm, using 10, 15 and 20 passes of \textsc{SysRem}. Currently there exists no reliable method of determining the optimal number of \textsc{SysRem} iterations beyond empirically optimising the strength of a recovered signal or injection. Though this method was used in \citet{Merritt2020}, it is not particularly feasible here given our search for a multitude of different species, and we instead choose to test a small range. \blindcorrection{As many of the species searched for are expected to exist in the stellar spectrum, this range encompasses a higher number of \textsc{SysRem} iterations than the 9 used in \citet{Merritt2020} in order to maximise removal of stellar residuals.}

We then cross-correlate each frame of each order with the corresponding section of the model spectra, binned down to the instrumental resolution of the order. \blindcorrection{As described in \citet{Merritt2020}, no planetary signal is expected in the out-of-transit frames. A basic transit model was thus generated using the equations of \citet{MandelAgol2002} (assuming no limb-darkening), and this was used to weight the time series of cross-correlation functions.}

\blindcorrection{\blindcorrection{A peak is seen in the cross-correlation function where a match is found between the template spectrum and the spectral residuals. Where such a peak is present, it will be Doppler-shifted according to the motion of the planet around the parent star.} This results in a diagonal trace in the cross-correlation time-series which follows the Doppler shift over time. This trace can often be seen with the naked eye in cases where the signal is strong \citep[e.g.][]{Snellen2010}. To enhance weaker signals, the cross-correlation time series is integrated over a} \blindcorrection{range of potential radial velocity curves. The radial velocity $v_{\text{p}}$ at each frame of the observations is given by:}
\[
v_{\text{p}}(\phi) = v_{\text{sys}} + K_\textrm{p}\sin{(2\pi\phi)},
\]
\noindent where $v_{\text{sys}}$ is the systemic velocity, $K_\textrm{p}$ is the radial velocity semi-amplitude of the planet, and $\phi$ is the orbital phase of the planet\blindcorrection{at the time of observation}, where $\phi$ = 0 represents the mid-transit time. \blindcorrection{As the spectra have already been corrected to the stellar rest frame, $v_{\text{sys}}$ is expected to be zero. The x-axis of the cross-correlation time series can therefore be considered to represent the difference in velocity -- or `lag' -- from the stellar rest frame.}

\blindcorrection{The cross-correlation time-series is then integrated over a range of potential values of $K_\textrm{p}$ by linearly interpolating each order to a radial velocity curve generated using values for $K_\textrm{p}$ from 100 to 400 $\text{km s}^{-1}$ in steps of \blindcorrection{1} $\text{km s}^{-1}$. This step-size is smaller than the average single resolution element of the original spectra} ($\sim$ 3.15 $\text{km s}^{-1}$ averaged over both arms). A smaller range than in \citet{Merritt2020} was chosen in order to speed up computational time, while still remaining large enough to judge the overall noise profile of the resultant map. \blindcorrection{For a subset of species, we extended the $K_\textrm{p}$ range to check for artefacts around $K_\textrm{p}=0$ and $v_{\text{sys}}=0$, which could arise from poorly-subtracted systematic effects. However, we found no such evidence.}

\blindcorrection{Each 2D order of $K_\textrm{p}$-interpolated CCFs is then summed over the column-wise time direction. These 1D sums are stacked in $K_\textrm{p}$ order, creating one 2D velocity heat-map for each spectral order. The y-axis of these heat maps is $K_\textrm{p}$; the x-axis $v_{\text{sys}}$ represents, as before, the velocity difference from the stellar rest-frame.} Examples of the cross-correlation function times series and the resulting velocity maps are shown for both the red and blue arms in the right column of Fig.~\ref{fig:prepro}.

\blindcorrection{The $v_{\text{sys}}$ scale of the cross-correlation functions, and thus the velocity heat-maps, is set by the wavelength scale of the spectral order. Therefore, due} to the varying resolution across the spectral range, the map for each order has a slightly different scale in $v_{\text{sys}}$. The maps for the blue and red arms are thus \blindcorrection{linearly} interpolated to a velocity scale made up of the mean of the $v_{\text{sys}}$ scales for that arm and summed over the orders to form a final map for the blue and the red arm. The individual order maps are also combined into a final red + blue map by interpolating to the mean of all $v_{\text{sys}}$ scales and summing again over the orders, which have already been optimally weighted by their variance: this should have the effect of strengthening extant signals by using the full spectral range of our data.

\blindcorrection{Due to our earlier removal of the continuum from the data, and the subsequent use of cross-correlation, the z-axis of the velocity heat-maps is in arbitrary units. To set a detection significance} we initially used the common method also used in \citet{Merritt2020}. \blindcorrection{An area from $-50$ to $50$ $\text{km s}^{-1}$ in $v_{\text{sys}}$ is masked in order to exclude any potential signal, and the standard deviation of the rest of the map calculated. The full unmasked map is then divided by this standard deviation.}

Again similarly to \citet{Merritt2020}, we also \blindcorrection{attempted obtaining detection significances using the more robust phase-shuffling method, as seen in \citet{Esteves2017}.} The time-series of cross-correlation functions is \blindcorrection{linearly} interpolated to a common $v_{\text{sys}}$ scale and summed over the orders to form a `master' set of CCFs, which is then shuffled randomly in time before regenerating the cross-correlation maps. This is performed 1000 times, and \blindcorrection{the standard deviation at each pixel of the maps found, creating a standard deviation map by which the original, unshuffled map was divided. This method} was only used on a limited subset of the maps (those in which potential detections were found) due to the computationally-intensive nature of the method. Also, as in \citet{Merritt2020}, we found no discernible difference using this method beyond a slight increase in detection significance for most signals. Once again, we prefer to use the method that generates a more conservative estimate of detection significance. Detection significance was, in all cases, set independently for the red, blue, and red + blue maps.

The fully-automated cross-correlation process was performed for every atomic model template  outlined in Sec.~\ref{sec:models}, with three cross-correlation maps (the blue arm, the red arm, and the combined red + blue map) generated and plotted for each model. The automated signal detection looked for the strongest peak within a range of 10 $\text{km s}^{-1}$ in $v_{\text{sys}}$ and 20 $\text{km s}^{-1}$ in $K_\textrm{p}$ from the `expected' signal position of 217 $\text{km s}^{-1}$ (from parameters in \citealt{Delrez2016}) and 0 $\text{km s}^{-1}$ (i.e. the stellar rest frame). Peaks in the maps of over 3\,$\sigma$ in this range were highlighted for further inspection, although in practice every cross-correlation map was investigated by eye for potential signals. \blindcorrection{Uncertainties were estimated for the measurements of $v_{\text{sys}}$ and $K_\textrm{p}$ by taking the full-width half-maximum (FWHM) of the signal in the relevant axis on the map and converting this to a standard deviation. This in effect assumes that the cross-correlation signal is proportional to the likelihood function, and that the uncertainties are uncorrelated with other parameters. A more detailed treatment of the uncertainties in the velocity constraints will be left to future work.}

\begin{table*}
\centering
\bgroup
\def\arraystretch{1.2}
\begin{tabular}[width=\textwidth]{ccccccc}
\hline
Species & $\sigma$ & $K_\textrm{p}$ ($\text{km s}^{-1}$) & $v_{\text{sys}}$ ($\text{km s}^{-1}$) & T (K) & $P_{\text{cloud}}$ (bar) & \textsc{SysRem} iterations \\ \hline \hline
\ion{Fe}{I}           & 8.8 & $200\pm41.1$ & $-7.3\pm5.8$  & 3000 & 0.05                     & 10 \\
\ion{Ca}{II}          & 6.4 & $235\pm54.4$ & $-7.3\pm11$  & 4500 & 1                        & 10 \\
\ion{Cr}{I}           & 5.0 & $198\pm75.6$ & $-5.5\pm4.6$  & 2000 & 1                        & 10 \\
\ion{V}{I}            & 4.4 & $226\pm44.2$ & $-5.5\pm7.0$  & 2000 & $5 \times 10^{-5}$       & 10 \\
\ion{Ca}{I}           & 4.2 & $199\pm24.6$ & $-5.5\pm3.5$  & 3000 & $5 \times 10^{-5}$       & 10 \\
\ion{H}{I}            & 5.7 & $236\pm82.4$ & $-4.6\pm12$  & 7000 & 1                        & 15 \\
\ion{K}{I}            & 4.4 & $198\pm21.2$ & $-6.4\pm2.7$  & 3000 & $5 \times 10^{-5}$       & 20 \\
\ion{Sc}{II}          & 4.2 & $212\pm26.3$ & $5.5\pm5.8$   & 2000 & $5 \times 10^{-5}$       & 20 \\
\ion{Mn}{I}           & 4.0 & $217\pm28.0$ & $-4.6\pm3.9$  & 2000 & 0.05                     & 10 \\
\ion{Co}{II}          & 3.8 & $218\pm15.7$ & $-7.3\pm1.9$  & 3000 & 1                        & 15 \\
\ion{Ni}{I}           & 3.7 & $221\pm9.3$ & $-11\pm1.6$   & 2500 & 0.05                     & 20 \\
\ion{Co}{I}           & 3.6 & $198\pm29.3$ & $-13\pm3.5$   & 2000 & 0.05                     & 10 \\
\ion{Cu}{I}           & 4.6 & $203\pm11.0$ & $-11\pm2.3$   & 3000 & $5 \times 10^{-5}$       & 10 \\
\ion{Sr}{I}           & 4.0 & $203\pm18.7$ & $0.9\pm1.9$   & 2500 & 0.001                    & 20 \\
\ion{V}{II}           & 3.7 & $210\pm8.1$ & $-11\pm1.6$   & 2000 & 1                        & 20 \\
\ion{Ti}{I}$^{\star}$ & 3.6 & $234\pm27.2$ & $-5.1\pm1.7$  & 2000 & $5 \times 10^{-5}$       & 10 \\
\ion{Sr}{II}          & 3.6 & $201\pm18.7$ & $-0.9\pm8.5$  & 5000 & $5 \times 10^{-5}$       & 10 \\
\ion{Ti}{II}          & 3.4 & $204\pm17.0$ & $-14\pm3.5$   & 7000 & 0.05                     & 10 \\
\ion{Fe}{II}          & 3.2 & $217\pm9.8$ & $-9.2\pm1.9$  & 3000 & 1                        & 10 \\
\ion{Mg}{I}           & 2.9 & $198\pm14.0$ & $-7.3\pm2.7$  & 3000 & 1                        & 15 \\
\hline
$^{\star}$ blue arm only &&&&&&\\
&&&&&&\\
\end{tabular}
\egroup
\\
\caption{Parameters for detected signals presented in this work, in order of overall confidence. Here $\sigma$, $K_\textrm{p}$ and $v_{\text{sys}}$ are the signal detection significance and velocity location from the combined red + blue maps (apart from \ion{Ti}{i}, for which the blue arm results are presented). \textbf{Uncertainties for these quantities were estimated from the FWHM of the signal, converted to a standard deviation.} T and $P_{\text{cloud}}$ are the temperature and cloud deck pressure level parameters of the model for which the strongest signal was retrieved for the species in question (shown in Fig.~\ref{fig:allmodels}). The \textsc{SysRem} iterations column provides the number of iterations which resulted in the strongest signal for the species.}
\label{table:detection}
\end{table*}

\begin{table*}
\centering
\bgroup
\def\arraystretch{1.2}
\begin{tabular}{cccccccc}
\hline
Species &
  \begin{tabular}[c]{@{}c@{}}Significance\\ $>3.5 \sigma$\end{tabular} &
  \begin{tabular}[c]{@{}c@{}}Expected \\ position\end{tabular} &
  \begin{tabular}[c]{@{}c@{}}Present in\\ injection\end{tabular} &
  \begin{tabular}[c]{@{}c@{}}Minimal noise \\ in map\end{tabular} &
  \begin{tabular}[c]{@{}c@{}}Detected in\\ WASP-121b?\end{tabular} &
  \begin{tabular}[c]{@{}c@{}}Detected in\\ other HJ?\end{tabular} &
  \begin{tabular}[c]{@{}c@{}}Strength of\\ \textbf{signal}\end{tabular} \\ \hline \hline
\ion{Fe}{I}           & \checkmark            & \checkmark & \checkmark & \checkmark & [1, 2, 3, 4, 5, 6]     & [A, B, C, D]      & Strong     \\
\ion{Ca}{II}          & \checkmark            & \checkmark & \checkmark & \checkmark & [6]                  & [E, F, G]         & Strong     \\
\ion{Cr}{I}           & \checkmark            & \checkmark & \checkmark & \checkmark & [2, 3, 6]              & [H]               & Strong     \\
\ion{V}{I}            & \checkmark            & \checkmark & \checkmark & \checkmark & [2, 3, 6]              & --                & Strong     \\
\ion{Ca}{I}           & \checkmark            & \checkmark & \checkmark & \checkmark & [3]                 & [I]               & Strong     \\
\ion{H}{I}            & \checkmark            & \checkmark & $\times$ & \checkmark & [4, 6]                 & [F, G]            & Tentative     \\
\ion{K}{I}            & \checkmark            & \checkmark & $\times$ & \checkmark & [6]                  & [J, K, L]         & Tentative  \\
\ion{Sc}{II}          & \checkmark            & $\times$ & \checkmark & \checkmark & --                  & [H]               & Tentative  \\
\ion{Mn}{I}           & \checkmark            & \checkmark & \checkmark & $\times$ & --                  & --                & Weak       \\
\ion{Co}{II}          & \checkmark            & \checkmark & \checkmark & $\times$ & --                  & --                & Weak       \\
\ion{Ni}{I}           & \checkmark            & $\times$ & \checkmark & $\times$ & [3]                 & --                & Weak       \\
\ion{Co}{I}           & \checkmark            & $\times$ & \checkmark & $\times$ & --                  & [H]               & Weak       \\
\ion{Cu}{I}           & \checkmark            & $\times$ & $\times$ & \checkmark & --                  & --                & Very weak  \\
\ion{Sr}{I}           & \checkmark            & $\times$ & \checkmark & $\times$ & --                  & --                & Very weak  \\
\ion{V}{II}           & \checkmark            & $\times$ & \checkmark & $\times$ & --                  & --                & Very weak  \\
\ion{Ti}{I}           & $\times^{\star}$    & \checkmark & \checkmark & $\times$ & --                  & --                & Very weak  \\
\ion{Sr}{II}          & \checkmark            & $\times$ & $\times$ & $\times$ & --                  & [H]               & Very weak  \\
\ion{Ti}{II}          & $\times$            & $\times$ & \checkmark & $\times$ & --                  & [D, H]            & \textbf{Insignificant}  \\
\ion{Fe}{II}          & $\times$            & $\times$ & \checkmark & $\times$ & [2, 6, 7]              & [C, D, F, H]      & \textbf{Insignificant}  \\
\ion{Mg}{I}           & $\times$            & \checkmark & $\times$ & $\times$ & [3, 6]                 & [H, M]            & \textbf{Insignificant}  \\
\hline
\multicolumn{8}{l}{$^{\star}$ significance $>$ 3.5 in blue arm only}
\cr\\
\end{tabular}
\egroup
\\
\caption{Detected signals and the detection criteria they fulfil from Sec.~\ref{sec:detcrit}. References for WASP-121b: 1) \citet{Gibson2020}, 2) \citet{Ben-Yami2020}, 3) \citet{Hoeijmakers2020}, 4) \citet{Cabot2020}, 5) \citet{Bourrier2020}, 6) \citet{Borsa2021}, 7) \citet{Sing2019}. References for other hot Jupiters: A) \citet{Nugroho2020_w33b}, B) \citet{Ehrenreich2020}, C) \citet{Stangret2020}, D) \citet{Hoeijmakers2018}, E) \citet{Nugroho2020_k20b}, F) \citet{CasasayasBarris2019}, G) \citet{Turner2020}, H) \citet{Hoeijmakers2019}, I) \citet{Astudillo-Defru2013}, J) \citet{Sing2011}, K) \citet{Colon2012}, L) \citet{Sedaghati2016}, M) \citet{Vidal-Madjar2013}. Citations for species found in other hot Jupiters are non-exhaustive.}
\label{table:criteria}
\end{table*}

\subsection{Injection tests}
\label{sec:inj}
In order to test our sensitivity to the species we are searching for, and to judge the reliability of any potential signals, we also performed injection tests for every model used in the signal retrieval process. Every model was convolved with the instrumental resolution using a top-hat function and Doppler shifted to conform to a $K_\textrm{p}$ curve of -217 $\text{km s}^{-1}$. \blindcorrection{This value is the negative of the expected $K_\textrm{p}$, and was selected as it leads to a reversed but otherwise identical velocity curve to the expected signal, thus ensuring maximum similarity to a real signal while also ensuring that the injection is not artificially boosted by an extant signal.}  We also take into account the fact that the red arm spectra are not corrected to the stellar rest-frame until after \textsc{SysRem} detrending by introducing a `de-correction' for barycentric and systemic velocity into the red-arm injected models. The Doppler-shifted models are then multiplied into the original extracted spectra before the pre-processing steps outlined in Sec.~\ref{sec:prepro}, after which the cross-correlation process continues almost identically. However, injection tests were performed with 15 passes of \textsc{SysRem} exclusively, due to the computationally-intensive nature of the process.

We also performed additional injection tests for all ionised species that showed signals in detection. The models generated in Sec.~\ref{sec:models} assume hydrostatic equilibrium, but from \citet{Sing2019} we are aware that WASP-121b is likely to have an escaping exosphere containing ionised species, resulting in much deeper spectral features than would be expected from our simple model. These injection tests used models generated in an identical manner to those described in Sec.~\ref{sec:models} for the higher temperature range of 3500 -- 9500 K. These models were then scaled so that their features matched the rough height of $\sim$ 0.25 -- 0.3 $R_p / R_\star$ reported for \ion{Fe}{ii} and \ion{Mg}{ii} in \citet{Sing2019}. For similar reasons, we also scaled the \ion{H}{i} models so that the H-$\alpha$ line matched the height of $\sim$ 0.2 $R_p / R_\star$ reported by \citet{Cabot2020}.

\begin{figure*}
	\includegraphics[width=\textwidth]{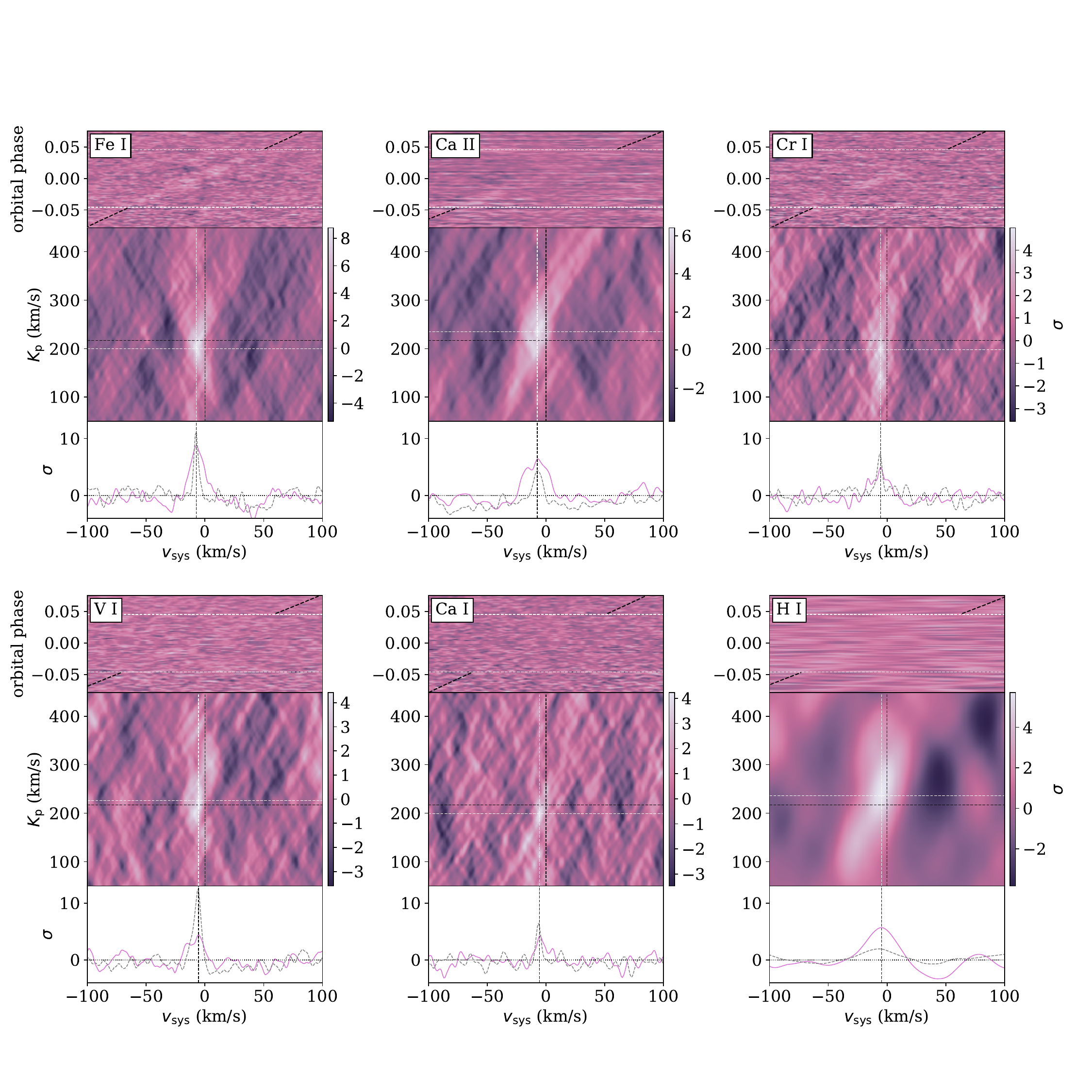}
    \caption{Results from cross-correlation for \ion{Fe}{i}, \ion{Ca}{ii}, \ion{Cr}{i}, \ion{V}{i}, \ion{Ca}{i} and \ion{H}{i}. Top: the summed CCFs for the blue and red arms. The dashed white lines indicate ingress and egress. The dashed black line indicates the expected position of the signal trace. Middle: the velocity map. The black dashed cross-hairs indicate the expected position of the signal, while the white cross-hairs centre on the peak of the detected signal. Bottom: a slice of the velocity map at the value of $K_\textrm{p}$ for which the signal was located. The purple line is from the velocity map presented: the dotted grey line shows the signal from the injection test of the same model, adjusted to have the same $v_{\text{sys}}$ offset as the detected signal. \blindcorrection{The smoother CCFs and map seen for \ion{H}{i} are the result of the template spectrum being dominated by a few broad lines.} The results for \ion{Ca}{ii} are presented in further detail in Fig.~\ref{fig:CaIIresults}.}
    \label{fig:majorresults}
\end{figure*}

\begin{figure*}
	\includegraphics[width=\textwidth]{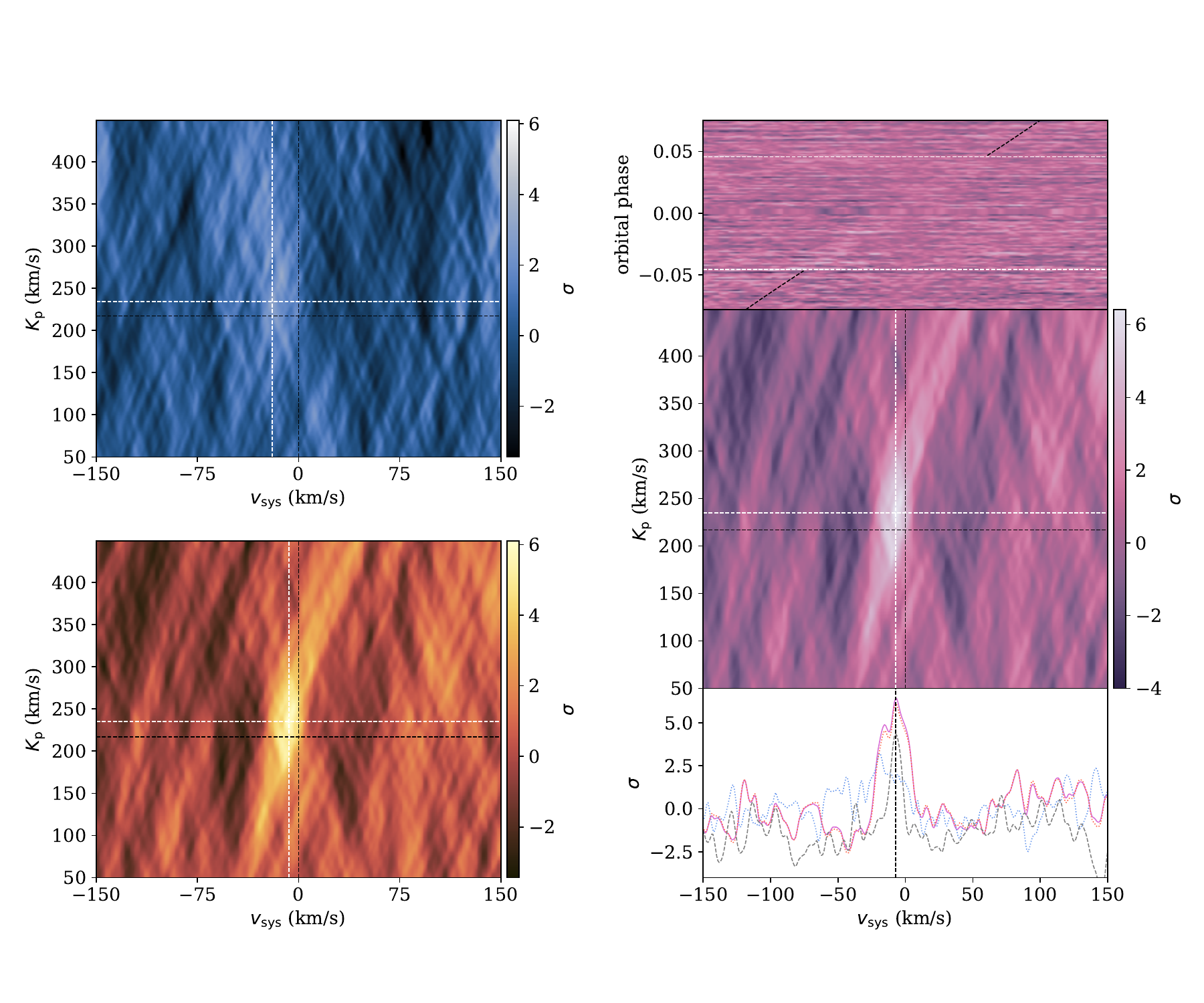}
    \caption{Results from cross-correlation for \ion{Ca}{ii}. Left: cross-correlation velocity maps for the blue arm (top) and the red arm (bottom), showing lack of signal in the blue arm. Right: as results presented in Fig.~\ref{fig:majorresults}. The red and blue dotted lines in the bottom-right plot represent slices through the red and blue maps at the detected $K_\textrm{p}$. The grey dashed line represents the injected signal for the same model with the line strength boosted to $\sim$ 0.25 -- 0.3 $R_p / R_\star$ and adjusted to have the same $v_{\text{sys}}$ offset as the detected signal.}
    \label{fig:CaIIresults}
\end{figure*}

\begin{figure*}
	\includegraphics[width=\textwidth]{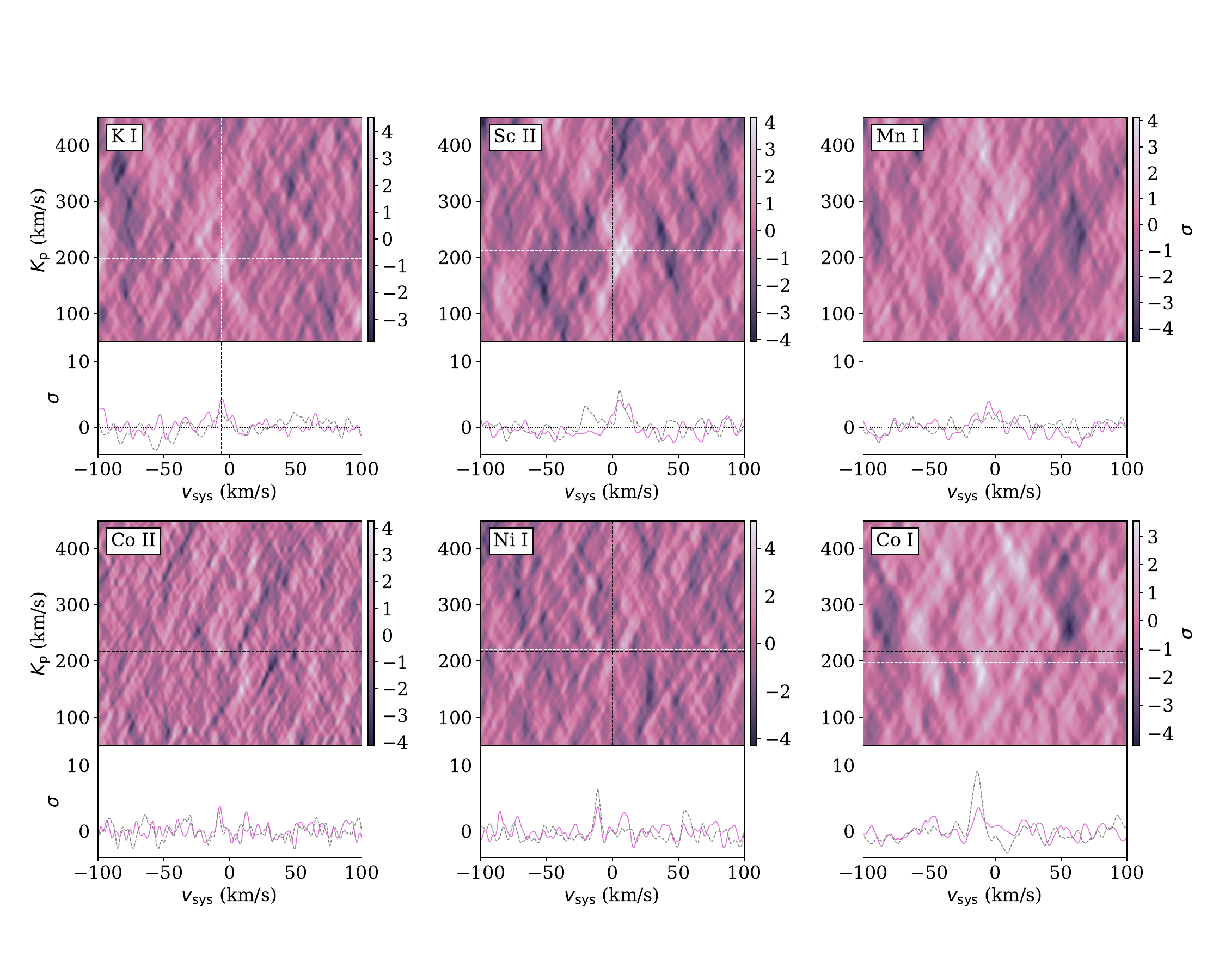}
    \caption{As Fig.~\ref{fig:majorresults}, for \ion{K}{i}, \ion{Sc}{ii}, \ion{Mn}{i}, \ion{Co}{ii}, \ion{Ni}{i} and \ion{Co}{i}. Note that the summed CCFs have been omitted from this figure due to the lack of visible trace.}
    \label{fig:mediumresults}
\end{figure*}

\section{Results}
\label{sec:res}

\subsection{Detection criteria}
\label{sec:detcrit}
The cross-correlation process resulted in a signal present within the detection range for \blindcorrection{17} different species, with a huge range in detection strength, signal position, and level of noise/additional structure in the cross-correlation map. \blindcorrection{Additionally, a further 3 signals were found outside of our detection range which are nonetheless of interest.} 

Simply applying a detection significance cut-off of 4\,$\sigma$, similarly to \citet{Merritt2020}, was found to be insufficient as a means of judging the reliability of the potential signals. For example, some otherwise-dubious signals were found to be stronger than 4\,$\sigma$, while some more likely signals fell just below. In addition, the methods used to generate the detection significances are, though common in the literature, rather simplistic, and rely on an assumption of Gaussian white-noise statistics which largely ignores the presence of structure in the map. In reality, the statistics of these cross-correlation maps are poorly understood, and the detection significance can be changed by as much as 0.5\,$\sigma$ by simply changing the area used to calculate the standard deviation, or by increasing/decreasing the number of phase-shuffling iterations. New techniques are being pursued and have produced encouraging results \citep{BrogiLine2019, Fisher2020, Gibson2020, Nugroho2020_w33b}, but due to their computationally-intensive nature, these methods were not attempted here for our large range of species.

As a result, we categorise our prospective detections using a more extensive set of diagnostic criteria. \blindcorrection{The intention of these criteria is primarily to rule out or weaken confidence in signals which may initially seem strong but are problematic in other areas. Signals must fulfill the majority of these criteria to be considered as detections.} The initial set of \blindcorrection{17} species was selected via examination by eye of all of the resulting cross-correlation maps. We then judged them considering the following criteria:

\begin{enumerate}

\item Detection strength at or above 3.5\,$\sigma$. Though it has been shown (e.g. by \citealt{Cabot2019}) that noise fluctuations in the cross-correlation map can reach the 4\,$\sigma$ level, we use a slightly less conservative value to avoid missing potentially-interesting signals. \blindcorrection{Any signals found below this cut-off were generally disregarded, with three potentially-interesting exceptions.}

\item Position of the signal. \citet{Gibson2020} recovered a strong \ion{Fe}{i} signal at a $v_{\text{sys}}$ offset of $\sim$ 5 $\text{km s}^{-1}$ and a $K_\textrm{p}$ of $\sim$ 190 $\text{km s}^{-1}$. We find the \ion{Fe}{i} signal at $K_\textrm{p}$ = $200\pm41.1$ $\text{km s}^{-1}$, $v_{\text{sys}}$ = $-7.3\pm5.8$ $\text{km s}^{-1}$ (see Fig.~\ref{fig:majorresults}). The negative $v_{\text{sys}}$ offset is also consistent with those observed by \citet{Hoeijmakers2020}, \citet{Ben-Yami2020}, \citet{Bourrier2020} and \citet{Cabot2020} and is thus thought to be reliable. We thus judge any signal that lies within 3 $\text{km s}^{-1}$ in $v_{\text{sys}}$ and 20 $\text{km s}^{-1}$ of $K_\textrm{p}$ of the \ion{Fe}{i} signal presented in \citet{Gibson2020} to lie at the `expected' position.

\item Whether the signal is detected in our injection tests. If a species is found as both a standard detection in the data and in injection, this increases confidence in the detection. Conversely, if no signal is found in injection for a species that presents a detected signal, we might assume that we do not in fact have the sensitivity to detect the species in question. Also, as we inject the signals at a velocity far from the prospective exoplanetary signal, this criterion accounts for cases where a model with very few lines `aliases' similar lines found in other extant species, potentially leading to a spurious detection. (See \ion{Co}{i}, Sec.~\ref{sec:weaksignals}.)

\item The presence of structure in the maps. This was assessed by a visual inspection to determine whether any peaks of similar size to the signal existed within the cross-correlation map. 

\item Whether the species has been detected before, whether in other analyses of WASP-121b \citep[e.g.][]{Hoeijmakers2020, Cabot2020, Ben-Yami2020}, or in other ultra-hot Jupiters such as KELT-9b \citep{Hoeijmakers2018, Hoeijmakers2019} or KELT-20b \citep[e.g.][]{CasasayasBarris2019, Nugroho2020_k20b}. Though we elect to remain as model-agnostic as possible, and thus assign no detectability criteria based upon whether a species is expected in WASP-121b via models, the actual previous detection of a species in WASP-121b or indeed in other ultra-hot Jupiters lends credibility to a previous detection.

\end{enumerate}

\begin{figure*}
	\includegraphics[width=\textwidth]{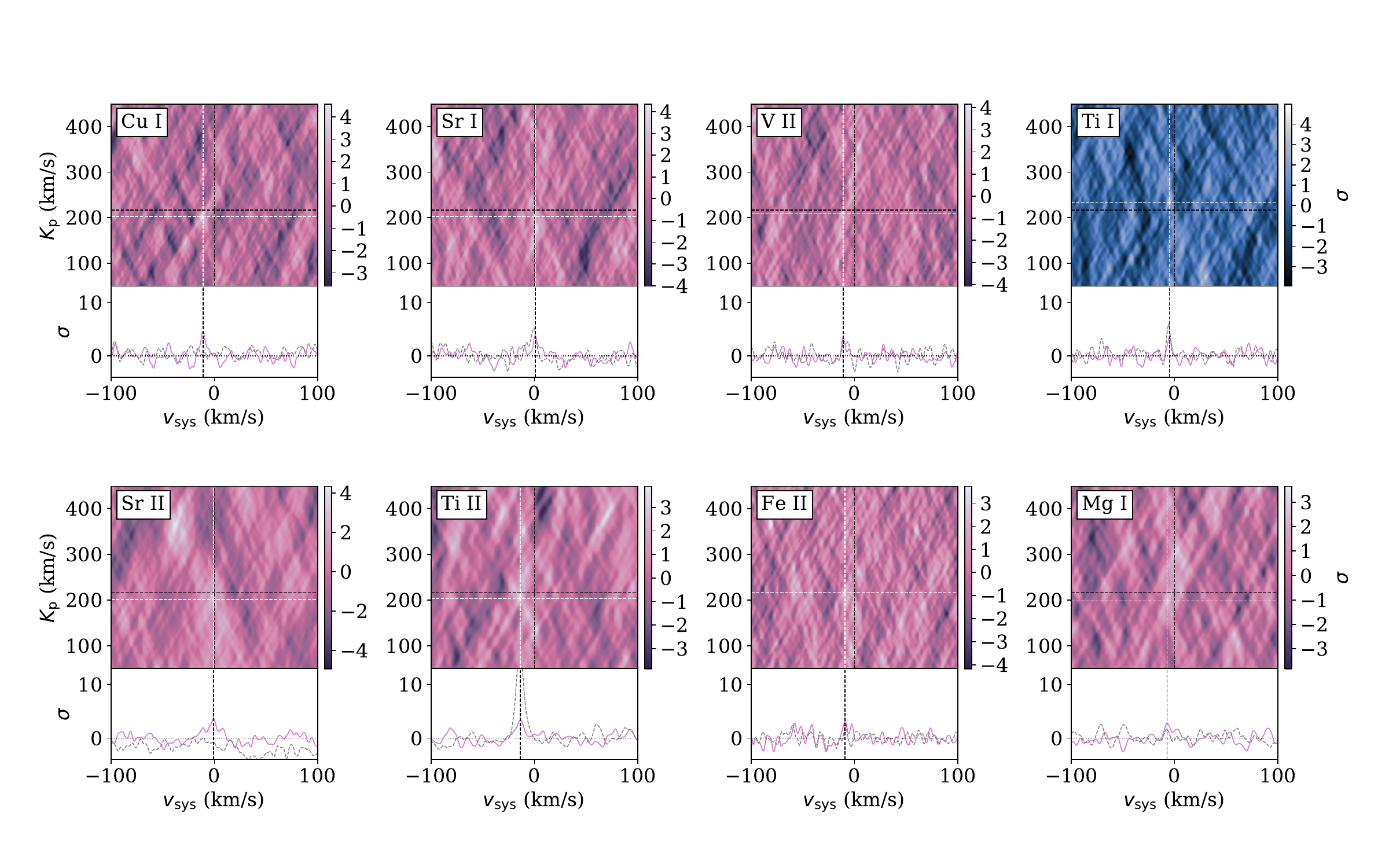}
    \caption{As Fig.~\ref{fig:mediumresults}, for \ion{Cu}{i}, \ion{Sr}{i}, \ion{V}{ii}, \ion{Ti}{i}, \ion{Sr}{ii}, \ion{Ti}{ii}, \ion{Fe}{ii}, and \ion{V}{ii}. Note that for \ion{Ti}{i}, the velocity map presented is from the blue arm only, as the signal did not appear in the combined blue + red velocity map.}
    \label{fig:minorresults}
\end{figure*}

\begin{figure*}
	\includegraphics[width=0.95\textwidth]{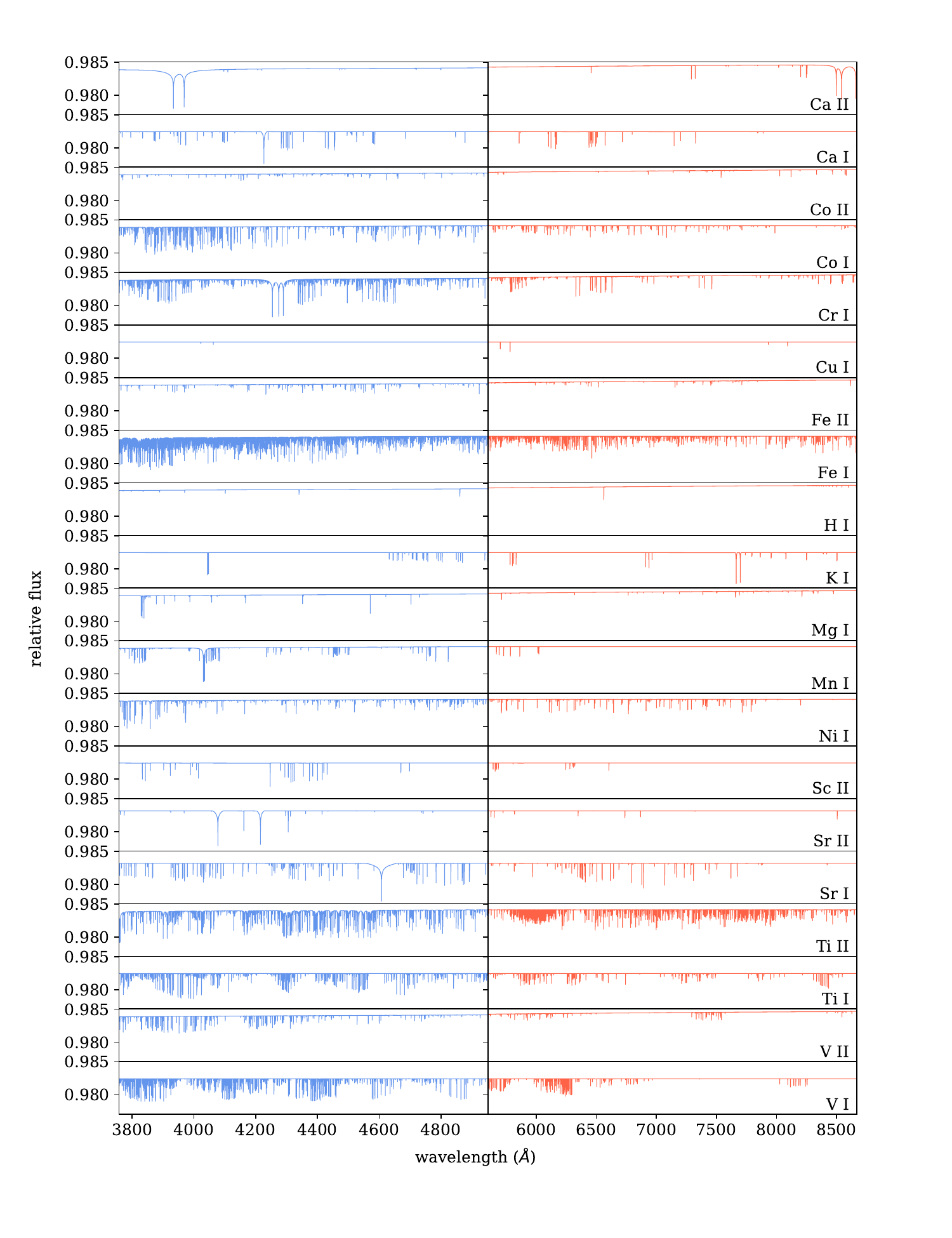}
	\centering
    \caption{The model transmission spectra used to recover the 20 signals presented in this work. The left column shows the wavelength range of the blue arm of our data: the right column shows the range of the red arm. Parameters of these models can be found in Table~\ref{table:detection}. All ion models are shown before the scaling described in Sec.~\ref{sec:inj}.}
    \label{fig:allmodels}
\end{figure*}

\subsection{Results of cross-correlation and injection tests}
A table of the 20 signals of interest and their parameters, sorted by the criteria fulfilled, is presented in Table~\ref{table:detection}. The detection criteria fulfilled for each species are elaborated upon on in Table~\ref{table:criteria}.

The cross-correlation maps and slices through the maps at the relevant $K_\textrm{p}$, with comparison to the injection tests, are found in Figs.~\ref{fig:majorresults}, \ref{fig:CaIIresults}, \ref{fig:mediumresults} and \ref{fig:minorresults}. In each case, we have chosen the result from model with the temperature and $P_{\text{cloud}}$ which provides the largest detection significance: these models can be found in Fig.~\ref{fig:allmodels}. Similarly, we present the results from the number of \textsc{SysRem} iterations that maximises the signal strength: however, we note that all signals are retrievable at 10 iterations. We reiterate here that the temperature and $P_{\text{cloud}}$ parameters are largely degenerate with other parameters; in each case, the `best' model is simply the one which provides the closest match to the exoplanetary signal out of those tested in terms of line strength above the continuum. The detection significances presented are those found using the more simplistic method of dividing the map by the standard deviation outside of the signal area. Phase-shuffling (as described in Sec.~\ref{sec:cc}) was found to vary the detection significance very slightly but otherwise produced very similar results, so we have reverted to the simpler, less computationally-intensive method here.

\blindcorrection{Only the eight signals found to fulfill four or five of our detection criteria are classed in this work as detections. The strongest and most reliable signals are of \ion{Fe}{i}, \ion{Cr}{i}, \ion{V}{i}, \ion{Ca}{i} and \ion{Ca}{ii}, fulfilling five out of five of the detection criteria; we classify these as as strong detections. Relatively interesting signals were also found of \ion{H}{i}, \ion{K}{i} and \ion{Sc}{ii}, fulfilling four criteria each, and are classified in this work as weak detections.}

\blindcorrection{Tentative signals were also found of \ion{Mn}{i}, \ion{Co}{ii}, \ion{Ni}{i} and \ion{Co}{i}: these signals were less reliable, fulfilling 3 of the detection criteria. Weaker hints were found of \ion{Cu}{i}, \ion{Sr}{i}, \ion{V}{ii}, \ion{Ti}{i} and \ion{Sr}{ii}. These satisfied only 1 -- 2 of our detection criteria. We do not claim any of these weaker signals as possible detections, and present them only to provide interesting avenues of exploration for future work or as some small additional evidence for already-discovered species. Finally, three species show signals below the detection threshold of 3.5\,$\sigma$: \ion{Fe}{ii}, \ion{Mg}{i}, and \ion{Ti}{ii}. We class these as technically insignificant but worthy of further discussion.} Further discussion of each of these signals in depth is presented in the following section. 

\subsection{Individual species results}
\label{sec:spec}
Given the large number of \blindcorrection{interesting signals, in the following section we discuss the eight signals classed as either strong or weak detections individually,} with reference to the cross-correlation maps shown in Figs.~\ref{fig:majorresults}, \ref{fig:CaIIresults}, and \ref{fig:mediumresults}. The weak signals are discussed as a whole in Sec.~\ref{sec:weaksignals} and are shown in Figs.~\ref{fig:mediumresults} and~\ref{fig:minorresults}.

\subsubsection{\ion{Fe}{i}}
As expected from previous work, we detect the \ion{Fe}{i} signal strongly at 8.8$\sigma$, within range of the expected location ($K_\textrm{p}$ = $200\pm41.1$ $\text{km s}^{-1}$, $v_{\text{sys}}$ = $-7.3\pm5.8$ $\text{km s}^{-1}$), in the combined blue + red map. This independent confirmation of our previous detection using both arms of the data fulfils all five of our detection criteria and serves as a benchmark test for our analysis, as the detection was previously well-characterised by \citet{Gibson2020} using the blue arm of this dataset and confirmed by \citet{Cabot2020}, \citet{Bourrier2020}, \citet{Ben-Yami2020} and \citet{Hoeijmakers2020} in the HARPS dataset. \blindcorrection{Our detection is consistent with these works.} Although we detect the \ion{Fe}{i} signal at a larger negative $v_{\text{sys}}$ offset than the values reported in the aforementioned works, we note that UVES is less stable in wavelength than HARPS, and less sensitive, with an average resolution element of $\sim 3.15~\text{km s}^{-1}$. In addition, this analysis was performed with an independent reduction and analysis pipeline.

\subsubsection{\ion{Ca}{ii}}
\label{sec:specCaII}
We retrieve a strong ($\sim 6.3 \sigma$) signal for \ion{Ca}{ii} in the expected location ($K_\textrm{p}$ = $236\pm54.4$ $\text{km s}^{-1}$, $v_{\text{sys}}$ = $-7.3\pm11$ $\text{km s}^{-1}$). As shown in Fig.~\ref{fig:CaIIresults}, we do not detect a significant signal in the blue arm of our data, which contains the strong \ion{Ca}{ii} H and K lines. We also did not detect a signal in the blue arm for any of our original \ion{Ca}{ii} injection tests. We theorise that this is due to the presence of extremely broad H and K lines in the stellar spectrum. Our data will have low S/N in the vicinity of these broad stellar lines, especially after optimal weighting. A similar effect was observed by \citet{Seidel2020} for sodium in the high-resolution transmission spectrum of the bloated super-Neptune WASP-166b. The majority of our detected signal comes instead from the strong triplet at the redward end of our red arm data.

Though \ion{Ca}{ii} was not found in our original injection tests using models limited by the assumption of hydrostatic equilibrium, it is recovered easily when we scale the injected models to account for a higher temperature and escaping atmosphere, as described in Sec.~\ref{sec:inj}. We also recover stronger signals at higher temperatures, with our strongest signal found using a model generated at 4500 K. We thus surmise, with reference to \citet{Sing2019} and \citet{Borsa2021}, that \ion{Ca}{ii} exists mainly in the hot extended exosphere of WASP-121b. This is discussed further in Sec.~\ref{sec:dis}. \ion{Ca}{ii} has previously been detected in the atmosphere of the UHJ KELT-20b/MASCARA-2b by \citet{CasasayasBarris2019} and \citet{Nugroho2020_k20b}, where it is also theorised to be part of an escaping atmosphere.

\subsubsection{\ion{Cr}{i} and \ion{V}{i}}
We confirm the detection of \ion{Cr}{i} by \citet{Ben-Yami2020}, \citet{Hoeijmakers2020} and \citet{Borsa2021}, recovering the signal at 5.0\,$\sigma$ at a position \blindcorrection{consistent with these works} ($K_\textrm{p}$ = $198+76$ $\text{km s}^{-1}$, $v_{\text{sys}}$ = $-5.5\pm4.6$ $\text{km s}^{-1}$) and fulfilling all five of our detection criteria. As can be seen in Fig~\ref{fig:prepro}, the signals are found in slightly different velocity locations in the blue and red arms. We discuss the variation in offsets in $v_{\text{sys}}$ further in Sec.~\ref{sec:dis}. 

We also confirm the detections of \ion{V}{i} \citep{Ben-Yami2020, Hoeijmakers2020, Borsa2021}, recovering the signal at 4.4\,$\sigma$ within range of the expected location ($K_\textrm{p}$ = $226\pm44.2$ $\text{km s}^{-1}$, $v_{\text{sys}}$ = $-5.5\pm7.0$ $\text{km s}^{-1}$) and fulfilling all five of our detection criteria. 

\subsubsection{\ion{Ca}{i}}
We retrieve a \ion{Ca}{i} signal of 4.2\,$\sigma$ close to the expected position, at $K_\textrm{p}$ = $199\pm24.6$ $\text{km s}^{-1}$ and $v_{\text{sys}}$ = $-5.5\pm3.5$ $\text{km s}^{-1}$. There is a small amount of structure in the cross-correlation map which, upon inspection of the maps from the individual arms, seems to result from an imperfect removal of the stellar \ion{Ca}{i} signal in the red arm. This imperfect removal may be due to the fact that our red arm data is detrended using \textsc{SysRem} in the rest frame of the Earth, in order to optimise the removal of the telluric spectrum, which is usually far more dominant within this wavelength range. Due to this structure in the cross-correlation map, this detection fulfils only four of our five detection criteria. However, \ion{Ca}{i} was detected in the HARPS datasets by \citet{Hoeijmakers2020}, giving further confidence in its detection in the UVES dataset.

\subsubsection{\ion{H}{i}}
We retrieve a strong but exceptionally diffuse signal for \ion{H}{i} at 5.7\,$\sigma$, at the expected location of $K_\textrm{p}$ = $236\pm82.4$ $\text{km s}^{-1}$ and $v_{\text{sys}}$ = $-4.6\pm12$ $\text{km s}^{-1}$. The diffuse nature of the signal can be attributed to the broad nature of the Balmer lines in both our model and in the signal present within the data. The high temperature at which we detect our strongest signal (7000 K) would support this hypothesis, as \ion{H}{i} is presumably, like \ion{Ca}{ii}, present in the extended atmosphere.  This broadening may also explain why the \ion{H}{i} signal is not present in our injection tests, even when the H-$\alpha$ line is scaled upwards to match the height of the feature found by \citet{Cabot2020} in WASP-121b using differential transit analysis.

\subsubsection{\ion{K}{i}}
The \ion{K}{i} signal is recovered at 4.4\,$\sigma$ and is found in the expected position, at $K_\textrm{p}$ = $198\pm21.2$ $\text{km s}^{-1}$ and $v_{\text{sys}}$ = $-6.42.7$ $\text{km s}^{-1}$; however, it is only present in injection tests at higher values of $P_{\text{cloud}}$ than the model we found to produce the strongest signal. Potassium is thought to be a commonly-detectable component of hot Jupiter atmospheres (\citealt{Fortney2010}, see also detections by e.g. \citealt{Sing2011}, \citealt{Colon2012}, \citealt{Sedaghati2016}), but it is expected to be largely ionised at the temperatures probed at the limb of WASP-121b. \blindcorrection{However, K was also detected in the analysis of ESPRESSO data of WASP-121b presented by \citet{Borsa2021}: our detection in the UVES data thus adds additional confidence to its presence.} It is possible that differential transit analysis focused on strong K lines could further confirm this detection \citep[see][for an example with UVES]{Gibson2019}. This method could additionally explore other species with few, strong lines in the wavelength region; previous detections made with this method in WASP-121b include \citet{Cabot2020}'s detection of both the \ion{Na}{i} doublet and an extended H$\alpha$ feature attributed to atmospheric escape.

\subsubsection{\ion{Sc}{ii}}
We retrieve a \ion{Sc}{ii} signal at 4.2\,$\sigma$. The signal seems to be fairly clear and well-defined within the cross-correlation maps, with a small amount of structure, and at $212\pm26.3$ $\text{km s}^{-1}$ is well-placed in $K_\textrm{p}$, but the position of the signal in systemic velocity is extremely unusual: the offset in $v_{\text{sys}}$ is positive, at $5.5\pm5.8$ $\text{km s}^{-1}$. \ion{Sc}{ii} has so far been discovered only in KELT-9b \citep{Hoeijmakers2019}, suggesting that if this signal is physical, we are probing a high-temperature region, perhaps in the exosphere (similarly to \ion{Ca}{ii}, Sec.~\ref{sec:specCaII}). This could also partially explain the very different $v_{\text{sys}}$. However, unlike \ion{Ca}{ii}, we do not retrieve the signal at a higher significance using higher-temperature models.  Due to this, and due to the unusual location of the signal, we encourage further investigation.

\begin{figure*}
	\includegraphics[width=0.95\textwidth]{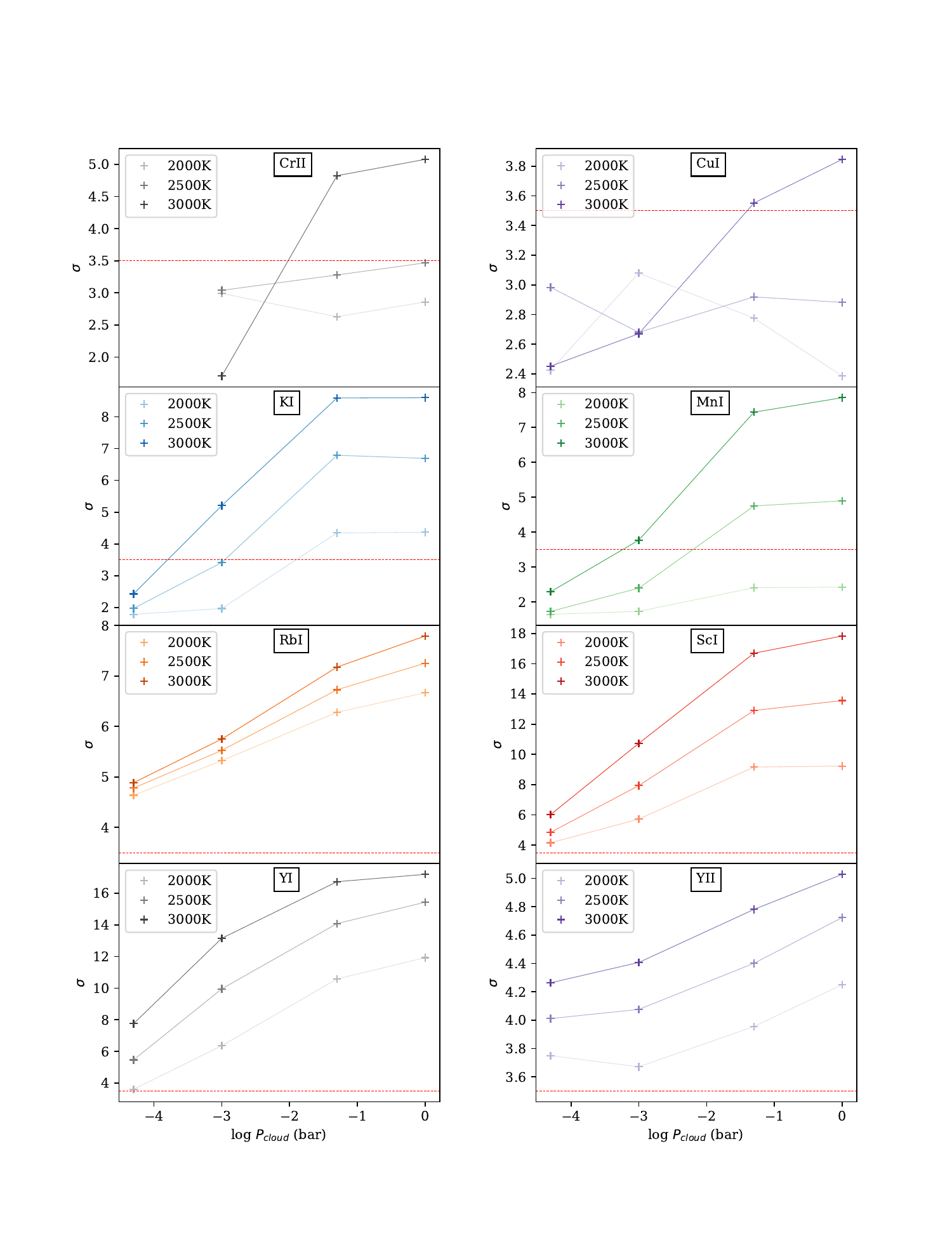}
    \caption{Results from injection tests for notable species, showing sigma vs. $P_{\text{cloud}}$ for every model injected. The red dotted line indicates\,$\sigma$ = 3.5, in accordance with the detection criteria in Sec.~\ref{sec:detcrit}.}
    \label{fig:inj_results}
\end{figure*}

\subsubsection{Other signals}
\label{sec:weaksignals}
\blindcorrection{While we classify none of the following as detections, we also find weak signals for \ion{Mn}{i}, \ion{Co}{ii}, \ion{Ni}{i} and \ion{Co}{i}; very weak signals for of \ion{Cu}{i}, \ion{Sr}{i}, \ion{V}{ii}, \ion{Ti}{i} and \ion{Sr}{ii}; and technically insignificant but interesting signals of \ion{Fe}{ii}, \ion{Mg}{i}, and \ion{Ti}{ii}.} Further details of these weaker signals can be found in Table~\ref{table:detection} and Figs.~\ref{fig:mediumresults} and~\ref{fig:minorresults}.

We find a signal for \ion{Mn}{i} at 4.0\,$\sigma$ and at the expected location. 
Though the signal appears quite clear, there is, however, a lot of structure in the map: the signal seems to have a `trail' pointing downwards in $K_\textrm{p}$, and there is an additional peak found at $K_\textrm{p}$ = $\sim$ 375 $\text{km s}^{-1}$. Additionally, the model for which we present the highest detection significance for \ion{Mn}{i} does not result in a detectable signal when injected, although \ion{Mn}{i} was found to be detectable in injection at higher temperatures/lower values for $P_{\text{cloud}}$ (see Fig.~\ref{fig:inj_results}). \blindcorrection{For these reasons, we do not claim it as a detection.} To date, \ion{Mn}{i} has not been confirmed in an exoplanet atmosphere. However, it was theorised by \citet{Lothringer2020} to be one of many species contributing to the increased transit depth at short wavelengths seen in low-resolution transmission spectra of UHJs, including WASP-121b \citep{Evans2018}.

A faint \ion{Co}{ii} signal is detected at 3.8\,$\sigma$, close to the expected location.
A \ion{Co}{ii} signal is also seen faintly in injection. However, the presence of structure in the map at a similar significance level greatly reduces our confidence in this signal, and \ion{Co}{ii} has not yet been detected in an exoplanet atmosphere. 

\blindcorrection{\ion{Mg}{i} and \ion{Ni}{i} have been previously detected by \citet{Hoeijmakers2020} in WASP-121b, with \ion{Mg}{i} also being detected by \citet{Borsa2021}. We do not recover either signal strongly. The \ion{Ni}{i} signal is detected at 3.7\,$\sigma$, with a large negative $v_{\text{sys}}$ offset and much structure in the CCF map. We detect only a very weak potential \ion{Mg}{i} signal, at 2.9\,$\sigma$. Our comparatively weaker signals for these species may be due to the higher resolution of HARPS and ESPRESSO at blue wavelengths and the wavelength coverage gap in our data between the red and blue arms. This gap, which is covered by the HARPS and ESPRESSO transits explored by \citet{Hoeijmakers2020} and \citet{Borsa2021}, contains a strong \ion{Mg}{i} triplet at $\sim$ 5200 \AA. We surmise that our lack of coverage of this triplet is responsible for the faintness of our \ion{Mg}{i} signal.} Though other strong \ion{Mg}{i} lines exist at the very blueward end of the blue arm of our data, this region of our spectra is low in S/N and is downweighted heavily in our analysis.

We retrieve a faint \ion{Co}{i} signal at 3.6\,$\sigma$. However, while its location in $K_\textrm{p}$ is within the expected range, the signal position is at a much greater negative offset in $v_{\text{sys}}$, at $-12.3\pm3.5$ $\text{km s}^{-1}$. A similarly large negative offset is found in several other potential signals, such as \ion{Ni}{i}, and is discussed further in Sec.~\ref{sec:dis}. The presence of similarly-sized structure in the map also reduces our confidence in this signal. Evidence of \ion{Co}{i} has previously been found in KELT-9b by \citet{Hoeijmakers2019}; \citet{Hoeijmakers2020} searched for \ion{Co}{i} in WASP-121b, but did not detect a signal.

The signal we recover for \ion{Cu}{i} is surprisingly high in significance (4.6\,$\sigma$), and presents a cross-correlation map largely free of similarly-scaled structure. While the $v_{\text{sys}}$ offset, at $-11\pm2.3$ $\text{km s}^{-1}$, is anomalous, it is no more so than several other signals retrieved by our methodology. However, examination of the model (found in Fig.~\ref{fig:allmodels}) shows that there are few lines present in the model over the wavelength range of our data. The signal appears to originate from just four lines in the red arm, none of which are particularly strong. Additionally, this species is not well-retrieved in our injection tests, especially not at the temperature and cloud-deck level used to recover the original signal. It is possible that these four lines are `aliasing' lines found in other extant species, producing a false positive. If this signal is spurious, as we suspect, then it justifies our use of a more elaborate set of detection criteria than simply the detection significance in\,$\sigma$ to judge the reliability of our detections. 

Both \ion{Sr}{i} and \ion{Sr}{ii} show faint signals, at 4.0 and 3.6\,$\sigma$ respectively. \ion{Sr}{ii} was previously tenuously detected by \citet{Hoeijmakers2019} in KELT-9b; similarly to their work, we also find a lot of positive structure in our cross-correlation map, casting doubt upon the veracity of the signals. We also recover a \ion{V}{ii} signal at 3.7\,$\sigma$, which if physical would indicate that enough of the \ion{V}{i} is being ionised to be detectable in transmission.

We also find a faint 3.2\,$\sigma$ signal for \ion{Fe}{ii}, but at a large negative $v_{\text{sys}}$ offset. \citet{Sing2019} found strong evidence of escaping \ion{Fe}{ii} in their NUV observations of WASP-121b, and an \ion{Fe}{ii} signal was recovered by both \citet{Ben-Yami2020} and \citet{Borsa2021}. However,  previous work on the blue arm of the dataset presented here by \citet{Gibson2020} did not retrieve a \ion{Fe}{ii} signal, and \citet{Hoeijmakers2020} could not repeat \citet{Ben-Yami2020}'s detection in the HARPS spectra. It is possible that variations in methodology are responsible for the elusive nature of this signal, or perhaps even some intrinsic variation in the signal itself caused by atmospheric dynamics. Alternatively, it is also possible that the putative \ion{Fe}{ii} signal found in cross-correlation here is simply an unfortunately-located noise peak, and our analysis is not sensitive to this species. It was suggested by \citet{Gibson2020} that the \ion{Fe}{ii} lines may be significantly broadened due to the larger velocity range within the escaping exosphere, leading to their removal during pre-processing and a resulting non-detection at high-resolution. \ion{Fe}{ii} is readily-detected in our injection tests at $\sim$ 12\,$\sigma$ when the injected model is scaled so the features match those found in \citet{Sing2019} (see Sec.~\ref{sec:inj}), suggesting that the low number of spectral lines in the model is not necessarily responsible for previous non-detections. A simple test in which we broadened the injected spectral lines with a Gaussian kernel did, however, reduce the strength of the recovered injected signal by as much as $\sim$ 5\,$\sigma$

Finally, we detect some small hints of both \ion{Ti}{i} and \ion{Ti}{ii}. \ion{Ti}{i} is seen only in the blue arm of the data at 3.6\,$\sigma$, close to the expected position. We attribute its lack of presence in the red+blue combined map to destructive addition with a negative noise peak seen in the red map. \ion{Ti}{ii} is seen in the combined red + blue map at 3.4\,$\sigma$. Both maps contain similarly-scaled structure, especially \ion{Ti}{ii}, which shows several peaks in the vicinity. \ion{Ti}{i} was searched for -- but not found -- in the HARPS dataset by both \citet{Ben-Yami2020} and \citet{Hoeijmakers2020}. \blindcorrection{\citet{Borsa2021} searched for both \ion{Ti}{i} and \ion{Ti}{ii} with no result.} We discuss the implications of this further in Sec.~\ref{sec:dis}.

\subsection{Injection tests and non-detections}
Of the 43 species we searched for in WASP-121b, we detect no sign of the remaining 23. Notably, our injection tests show that several species we did not detect in our broad species search are nevertheless theoretically detectable using our methodology for given values of T and $P_{\text{cloud}}$: \ion{Y}{i}, \ion{Y}{ii}, \ion{Rb}{i}, \ion{Sc}{i} and \ion{Cr}{ii}. Additionally, though the specific models used to retrieve the \ion{K}{i}, \ion{Cu}{i} and \ion{Mn}{i} signals do not present a signal in injection tests, these species are nonetheless detectable at higher values of T or lower values of $P_{\text{cloud}}$. We present the results from these injection tests in Fig.~\ref{fig:inj_results}. 

Unlike in \citet{Merritt2020}, we do not choose to set detection limits upon these species based upon the results of the injection tests, as here we chose to vary the position of the grey cloud deck $P_{\text{cloud}}$ rather than VMR. As discussed in \citet{Merritt2020}, even had we varied VMR, any detection limits we set on abundances would be extremely contingent on the correct positioning of the cloud deck due to the unbreakable degeneracies present using our simplified atmospheric model. 

\section{Discussion}
\label{sec:dis}
\blindcorrection{Our methodology resulted in five detections, three tentative detections and weak signs of a further 9 different atomic species} in the atmosphere of WASP-121b using a single UVES transit, standard cross-correlation methodology and model transmission spectra generated using a simple analytical approximation. Though we emphasise that \blindcorrection{only eight of these signals were strong enough to be classified as potential detections}, our success nevertheless shows the effectiveness of high-resolution broad species searches in UHJs. We provide an independent confirmation of many of the species detected in the HARPS dataset explored by \citet{Ben-Yami2020} and \citet{Hoeijmakers2020} \blindcorrection{and the ESPRESSO dataset explored in \citet{Borsa2021}. We also add a novel (if tentative) detection of \ion{Sc}{ii}.}

The presence alone of various species provides information on atmospheric conditions at the limb of WASP-121b. The existence of transition metals such as Fe, V and Cr in UHJ atmospheres was predicted by theoretical work using stellar models by \citet{Lothringer2018}, \citet{LothringerBarman2019} and \citet{Lothringer2020}, who posited that the strong optical opacity of these species would be sufficient to drive the thermal inversion observed in emission \citep{Evans2017, Mikal-Evans2019}. It has also been posited \citep[e.g., by][]{Lothringer2020} that a forest of transition metal lines could be responsible for the strong opacity seen at wavelengths $<$ 3000 \AA in transmission spectra of WASP-121b \citep{Evans2018}, WASP-76b \citep{Fu2020arXiv}, and WASP-12b \citep{Fossati2010, Haswell2012, Sing2013}. Due to the dense region of overlapping lines they create, these species are extremely difficult to isolate at low-resolution, highlighting the importance of high-resolution observations for confirming the presence of such species. Additionally, the presence of any atomic species indicates that it is not substantially condensed out at the limb of WASP-121b, indicating that substantial regions of the limb of WASP-121b probed by our analysis are $>$ 2000K: below this temperature, we expect atomic species to largely have condensed out of the atmosphere \citep{Hoeijmakers2020}.

Our strong detection of \ion{Ca}{ii}, tentative detection of \ion{H}{i}, and weaker signals for ions such as \ion{Sc}{ii}, \ion{Co}{ii}, \ion{Sc}{ii}, \ion{Sr}{ii}, \ion{Ti}{ii} and \ion{V}{ii}, indicates that our analysis is probing much higher temperatures than might be expected from previous studies of the limb of WASP-121b \citep[e.g.][]{Evans2018, Gibson2020}. Signs of \ion{Sc}{ii} and \ion{Sr}{ii} have previously only been detected in the extremely hot Jupiter KELT 9b (> 4000 K, \citealt{Hoeijmakers2018, Hoeijmakers2019}). We therefore infer that these detections arise from the existence of a hot escaping atmosphere, as detected by \citep{Sing2019}. This hypothesis is strengthened by the diffuse nature of our recovered \ion{Ca}{ii} and \ion{H}{i} signals, suggesting the presence of broadened spectral lines in the transmission spectrum, and by the fact that despite the strength of our detection, \ion{Ca}{ii} is only detected in injection tests when we `boost' the size of the spectral features in our models to $\sim$ 0.25 -- 0.3 $R_p / R_\star$, similar to those found by \citet{Sing2019} for \ion{Fe}{ii} and \ion{Mg}{ii}.

Our analysis also discovered hints of \ion{Ti}{i} and \ion{Ti}{ii}, neither of which were found in the HARPS dataset by \citet{Ben-Yami2020} and \citet{Hoeijmakers2020}. Though we emphasise that we have found only the slightest trace of a signal for both species, their potential presence is intriguing. As mentioned, WASP-121b is known to host a temperature inversion due to the detection of water features in emission \citep{Evans2017, Mikal-Evans2019, Mikal-Evans2020}, but so far no work to date has found convincing evidence of TiO, one of the molecules thought to be responsible for driving temperature in versions in the hottest of hot Jupiters \citep{Hubeny2003, Fortney2008}. Indeed, an explicit search for TiO in the red arm of this data by \citet{Merritt2020} using a similar methodology to that presented here found nothing. It has been suggested that perhaps absorption by atomic metals, most notably \ion{Fe}{i}, may be responsible for the atmospheric heating driving the temperature inversion \citep{Lothringer2018, Gibson2020}, and that the lack of Ti and TiO in the atmosphere is due to a `cold-trap' mechanism: the lower condensation temperature of TiO is causing it to condense out and become trapped in condensate form in cooler areas of the atmosphere \citep{Lodders2002, Hubeny2003, Spiegel2009, Parmentier2013, Parmentier2016, Beatty2017}. In this scenario, \citet{Hoeijmakers2020} posit that Ti would also be depleted, as chemical equilibrium would respond to the depletion of TiO by driving more and more atomic Ti into its oxide phase, causing further condensation until all Ti-bearing species are condensed out of their gas phase. A detection of \ion{Ti}{i} and \ion{Ti}{ii} would seem to contradict this explanation, and render the lack of TiO even more mysterious. However, we emphasise that our signals are weak, and hope for further exploration with larger data sets consisting of more transits, perhaps at higher resolution, in order to confirm or rule out the presence of neutral or ionised atomic titanium.

We see a large amount of variance in the position of our signals, in both $v_{\text{sys}}$ and $K_\textrm{p}$. The variation in $K_\textrm{p}$ is to some extent expected. In addition to the approximate $\pm$15 $\text{km s}^{-1}$ uncertainty on the `true' value of $K_\textrm{p}$ (calculated from parameters presented in \citealt{Delrez2016}), only a small portion of the planetary orbit is sampled during a transit, resulting in a large spread in $K_\textrm{p}$ \citep{Brogi2018}. Similar variation is seen in \citet{Hoeijmakers2020} and \citet{Ben-Yami2020}, with \citet{Ben-Yami2020} seeing variance of $\sim$ 50 $\text{km s}^{-1}$ in $K_\textrm{p}$.

The potential sources of variation in $v_{\text{sys}}$ are more complex. The slightly-differing signal location between red and blue arms seen in Fig.~\ref{fig:prepro} is a fairly common feature of the detected signals presented in this work. If physical, it may signify that we are probing different atmospheric regimes with the two different wavelength ranges. However, given the known instability of UVES in wavelength when compared with more stable instruments like HARPS, and the different methods used to correct the wavelength solution for the blue and red arms outlined in Sec.~\ref{sec:obs}, we suspect that this inter-arm variation in $v_{\text{sys}}$ is due to differences in alignment and wavelength solution accuracy between the blue and red arms. 

We also see large shifts in $v_{\text{sys}}$ in the combined red+blue maps for a large number of the signals we recover. As previously mentioned, \ion{Fe}{i} has been consistently recovered with a blue-shifted $v_{\text{sys}}$ offset of -3 -- -5 $\text{km s}^{-1}$ \citep{Gibson2020, Cabot2020, Bourrier2020, Ben-Yami2020, Hoeijmakers2020}, which has been attributed to the presence of atmospheric dynamics: namely, strong day-to-nightside winds, predicted to be on the order of $\approx$ 5 $\text{km s}^{-1}$ \citep[e.g.][]{Kataria2016}. Our \ion{Fe}{i} signal is found at -7.3 $\text{km s}^{-1}$, broadly in agreement with previous measurements, especially considering that UVES is less stable in wavelength than HARPS, and less sensitive, with an average resolution element of $\sim 3.15~\text{km s}^{-1}$. However, many of the more tentative signals shown in Fig.~\ref{fig:mediumresults} and Fig.~\ref{fig:minorresults} are found at more variant $v_{\text{sys}}$ offsets. For example, \ion{Co}{i}, \ion{Ni}{i}, \ion{Co}{ii} and several of the weakest signals are found at larger negative $v_{\text{sys}}$ offsets, on the order of $\approx$ - 10 $\text{km s}^{-1}$. If physical, it could perhaps be possible that these species are predominantly found in areas of the atmosphere for which the wind speed is even higher, leading to a larger blue-shift in the signal. However, winds this strong in the atmosphere are not predicted by current GCMs of UHJs \citep{Kataria2016}. Instead, it is possible that a variant blue-shift may be caused by material escaping the planet, pushed away from the planet and the star by stellar pressure. Hydrogen has been observed to reach velocities of up to 100 $\text{km s}^{-1}$ in the escaping tail of HD209458b \citep{Vidal-Madjar2003, Ben-Jaffel2007}. Speeds in the upper atmosphere could vary between species as a result of either differing atomic weight or ionisation state (due to interaction with magnetic fields).

Additionally, some species such as \ion{Sc}{ii} and \ion{Mg}{i} show a red-shift in $v_{\text{sys}}$. As our observations probe the entire annulus of WASP-121b during transit, any signals we retrieve are the result of an average over the limb. If, for example, clouds were blocking the side of the limb rotating towards us, this could result in an overall redshift being detected, though we note that it would seem very odd for this to happen to only some of the many species we have found evidence for. Significant cloud mass is also not expected in UHJs due to the high temperatures preventing condensates from forming, and this would be especially true for the hotter, blue-shifted evening terminator. Nevertheless, GCMs by \citet{Flowers2019} for the cooler hot Jupiter HD 189733b show that a combination of winds, rotation and clouds can lead to velocity shifts of $\pm$ 10 $\text{km s}^{-1}$ in different regions of the limb, so this explanation lies within the bounds of theoretical possibility. Also, \citet{Ehrenreich2020} resolve their detection of \ion{Fe}{i} in the atmosphere of the UHJ WASP-76b to have a blueshift of - 11 $\text{km s}^{-1}$ on the `evening' terminator and detect no signal from the nightside close to the morning terminator, which they attribute to the condensation of iron across the cooler nightside. A chemical gradient across the surface of the nightside of hot Jupiters is predicted by theory \citep[e.g.][]{Komacek2016, BellCowan2018} and could potentially explain both the large blue- and red-shifts seen in our data, depending on the exact chemistry involved. It is possible that using spectral models based on 3D atmospheric circulation models could result in stronger detections, as was found for emission in infrared by \citet{Beltz2020}, and help uncover the cause of these offsets.

However, we find it more likely that the source of this variance in $v_{\text{sys}}$ is simply due to the inherent wavelength instability of UVES. During the alignment process outlined in Sec.~\ref{sec:obs}, we see order-to-order variation in the wavelength solution for the blue arm of $\sim$ 3 $\text{km s}^{-1}$. While we attempt to correct this with alignment to a stellar spectrum, it is possible that some variation remains, and we additionally make no attempt to refine the dispersion in each order. Also, the red arm was not corrected for order-to-order variation due to both its smaller amplitude and the difficulty of aligning telluric-free orders to the telluric spectrum used for alignment. If, for example, some orders are more divergent from the correct wavelength solution than others, then species with the majority of their lines in these orders would naturally present signals with variant $v_{\text{sys}}$ offsets. This more prosaic explanation is supported by the fact that the our strongest signals are all detected at a $v_{\text{sys}}$ offset broadly consistent with previous works.

\section{Conclusions}
\label{sec:conc}
We have presented the results of a broad search for atomic species in the atmosphere of the ultra-hot Jupiter WASP-121b using high-resolution spectroscopy. Using standard cross-correlation methodology on a single transit observation taken with UVES, we recovered potential signals for \blindcorrection{17} neutral and ionised atomic species. Using five detection criteria, we confirm strong detections of \ion{Fe}{i}, \ion{Ca}{ii}, \ion{Cr}{i}, \ion{V}{i} and \ion{Ca}{i} and tentative detections of \ion{H}{i}, \ion{K}{i} and \ion{Sc}{ii}. We also uncover weak evidence for \ion{Mn}{i}, \ion{Co}{ii}, \ion{Ni}{i} and \ion{Co}{i}, and very weak hints of \ion{Cu}{i}, \ion{Sr}{i}, \ion{V}{ii}, \ion{Ti}{i} and \ion{Sr}{ii}. \blindcorrection{Technically insignificant yet potentially interesting signals of \ion{Fe}{ii}, \ion{Mg}{i}, and \ion{Ti}{ii} are also discussed.}

We have therefore presented independent confirmation or further evidence for previous detections of \ion{Cr}{i}, \ion{V}{i}, \ion{Ca}{i} and \ion{Ni}{i} made by \citet{Ben-Yami2020} and \citet{Hoeijmakers2020} using HARPS, \blindcorrection{and for the detections of \ion{K}{i} and exospheric \ion{Ca}{ii} made by \citet{Borsa2021} using ESPRESSO. We additionally presented evidence of exospheric \ion{H}{i}, previously found by by \citet{Cabot2020} using HARPS, and present a novel detection of \ion{Sc}{ii} at 4.2\,$\sigma$.} Finally, we have shown via injection tests that our methodology is sensitive to \ion{Y}{i}, \ion{Y}{ii}, \ion{Rb}{i}, \ion{Sc}{i} and \ion{Cr}{ii}, species for which we do not find significant evidence, though we decline to set detection limits upon these species due the degeneracies present. The detection of such a wide range of atomic species allows us to begin to set constraints on the temperature and refractory properties of WASP-121b, and provides a useful starting-point for more in-depth characterisation of the exoplanet atmosphere.

The success of our search, in both confirming previous detections and in recovering a large number of potential signals, echoes that of previous successful broad species searches made in high-resolution for UHJs such as KELT-9b \citep{Hoeijmakers2018, Hoeijmakers2019} and KELT-20b/MASCARA-2b \citep{CasasayasBarris2019, Nugroho2020_k20b}. Our recovery of a variety of signals in a single transit is highly encouraging, and implies that the potential of high-resolution spectroscopy for detecting atomic species in UHJs is as yet mostly untapped. However, although we have presented a large number of potential signals in this work, we emphasise that many of them are extremely weak or tentative. Future work using a greater number of transits, or with instruments with higher resolution and stability such as VLT/ESPRESSO, is encouraged to investigate the potential presence of many of the species presented herein, as well as to further investigate the velocity offsets of each detected species \citep[e.g.][]{Ehrenreich2020}. We have also refrained from placing any solid constraints on the abundances of our detected species, or on the atmospheric structure of WASP-121b. Our removal of the continuum prevents the measurement of any pressure-sensitive features which would allow the degeneracies between abundances, temperature, scattering properties and reference radius/pressure to be broken, and the cross-correlation method is not sensitive to changes in amplitude caused by differing scale heights. New and more complex approaches to high-resolution spectroscopic analysis have emerged in recent years, including the combination of low- and high-resolution spectroscopy \citep{Brogi2017, Pino2018}, principled statistical frameworks and likelihood mapping \citep{BrogiLine2019, Gibson2020, Nugroho2020_w33b, Hood2020}, machine learning \citep{Fisher2020} and Doppler tomography (\citealt{Watson2019}, Matthews et al. \textit{in prep}). These sophisticated methods enable more stringent constraints to be placed on atmospheric parameters, and broad species searches such as ours present a starting-point for future work using these sophisticated and more computationally-intensive methods to further categorise the atmosphere of WASP-121b.

\section*{Acknowledgements}

The authors are very grateful to the anonymous referee for their suggestions and observations. We would also like to thank Laura McKemmish for many useful comments. This work is based on observations collected at the European Organisation for Astronomical Research in the Southern Hemisphere under ESO programme 098.C-0547. N.P.G. gratefully acknowledges support from Science Foundation Ireland and the Royal Society in the form of a University Research Fellowship. S.K.N. and C.A.W. would like to acknowledge support from UK Science Technology and Facility Council grant ST/P000312/1. We are exceptionally grateful to the developers of the \textsc{NumPy}, \textsc{SciPy}, \textsc{Matplotlib}, \textsc{iPython}, \textsc{scikit-learn}, \textsc{AstroPy} and \textsc{numba} packages, which were used extensively in this work \citep{Harris2020, Virtanen2020, Hunter2007, PerezGranger2007, Pedregosa2012, Astropy2013, Lam2015}. The perceptually-uniform scientific colour maps used in Figs.~\ref{fig:prepro}, \ref{fig:majorresults}, \ref{fig:CaIIresults}, \ref{fig:mediumresults} and \ref{fig:minorresults} are by \citet{Crameri2018}.

\section*{Data Availability}
The observations underlying this analysis are publicly available in the ESO Science Archive Facility under program name 098.C-0547. Other data will be shared on reasonable request to the corresponding author.



\bibliographystyle{mnras}
\bibliography{bibs} 







\bsp	
\label{lastpage}
\end{document}